\documentclass[twocolumn,final,3p,times]{autart}
\usepackage[latin9]{inputenc}
\usepackage{float}
\usepackage{amsmath}
\usepackage{amssymb}
\usepackage{graphicx}

\makeatletter

\floatstyle{ruled}
\newfloat{algorithm}{tbp}{loa}
\providecommand{\algorithmname}{Algorithm}
\floatname{algorithm}{\protect\algorithmname}

\newtheorem{notation}[thm]{Notation}
\newtheorem{assumption}[thm]{Assumption}

\usepackage{algorithmic}

\usepackage{savesym}
\savesymbol{AND}

\journal{Automatica}

\makeatother

\begin{document}
\begin{frontmatter}

\title{Accuracy Analysis for Distributed Weighted Least-Squares Estimation in Finite Steps and Loopy Networks}

\author[dalian]{Tianju Sui}\ead{kevindulant@126.com}
\author[gua,arg]{Damián Marelli}\ead{Damian.Marelli@newcastle.edu.au}
\author[new,gua]{Minyue Fu}\ead{Minyue.Fu@newcastle.edu.au}
\author[gua]{Renquan Lu}\ead{rqlu@gdut.edu.cn}

\address[dalian]{Department of Control Science and Engineering, Dalian University of Technology, Liaoning, China.}
\address[gua]{School of Automation, Guandong University of Technology, Guangzhou, China.}
\address[arg]{French-Argentinean International Center for Information and Systems Sciences, National Scientific and Technical Research Council, Ocampo y Esmeralda, Rosario 2000, Argentina.}
\address[new]{School of Electrical Engineering and Computer Science, University of Newcastle, University Drive, Callaghan, NSW 2308, Australia.} 

\begin{abstract}
Distributed parameter estimation for large-scale systems is an active research problem. The goal is to derive a distributed algorithm in which each agent obtains a local estimate of its own subset of the global parameter vector, based on local measurements as well as information received from its neighbours. A recent algorithm has been proposed, which yields the optimal solution (i.e., the one that would be obtained using a centralized method) in finite time, provided the communication network forms an acyclic graph. If instead, the graph is cyclic, the only available alternative algorithm, which is based on iterative matrix inversion, achieving the optimal solution, does so asymptotically. However, it is also known that, in the cyclic case, the algorithm designed for acyclic graphs produces a solution which, although non optimal, is highly accurate. In this paper we do a theoretical study of the accuracy of this algorithm, in communication networks forming cyclic graphs. To this end, we provide bounds for the sub-optimality of the estimation error and the estimation error covariance, for a class of systems whose topological sparsity and signal-to-noise ratio satisfy certain condition. Our results show that, at each node, the accuracy improves exponentially with the so-called loop-free depth. Also, although the algorithm no longer converges in finite time in the case of cyclic graphs, simulation results show that the convergence is significantly faster than that of methods based on iterative matrix inversion. Our results suggest that, depending on the loop-free depth, the studied algorithm may be the preferred option even in applications with cyclic communication graphs.
\end{abstract}

\begin{keyword} Distributed statistical estimation, Weighted least squares, Convergence rate. \end{keyword}

\end{frontmatter}

\section{Introduction \label{sec:Introduction}}

With the fast development of sensor networks and wireless communications,
the scale of systems is becoming increasingly large. Since centralized
estimation requires a fusion center to process all the information
from the whole graph, the computation and communication burden increases
with the system's size. Thus, the centralized estimation approach
is not suitable for large-scale systems, and distributed approaches
are needed. The development of distributed estimation has attracted
a great deal of attention~\cite{Garin2010,li2009disenergy,gupta2009data,ribeiro2006a,ribeiro2006b}.
It finds applications in industrial monitoring, multi-agent systems,
the smart grid, etc.

The distributed estimation problem consists of a network of interconnected
nodes, each of which aims to obtain an estimate of certain vector
of interest. This is achieved through an iterative procedure in which
each node processes its available information, and exchange relevant
information with its neighbors, in order to successively compute the
required estimate as accurately as possible. The existing distributed
estimation problems can be broadly classified into four classes. These
classes are: static fully reconstructive, static partially reconstructive,
dynamic fully reconstructive and dynamic partially reconstructive.
A fully reconstructive system is one in which each node aims to obtain
an estimate of the same vector. In contrast, in a partially reconstructive
system, each node aims to obtain an estimate of its own partial sub-vector
of interest. Also, a static system is one in which prior knowledge
of the state at a certain time is independent of the knowledge of
the same state at previous times. A dynamic system refers to the complementary
case. We point out that methods for dynamic estimation can be readily
used for static problems, by choosing the dynamic model in a way such
that the state stays constant over time.

In the static fully reconstructive problem, the most popular distributed
estimation algorithm is consensus~\cite{Garin2010}. By running average
consensus on the information vector and information matrix of each
node, in view of the weighted least squares (WLS) formula, the final
estimate of each node converges to the one obtained via WLS~\cite{Olfati2005}.
Although the average consensus algorithm is simple, it has two main
disadvantages: First, the communication burden is large, as each node
communicates $\frac{n\times(n+3)}{2}$ scalars to its neighbors, where
$n$ is the dimension of the estimated vector. Second, the convergence
of average consensus requires infinite iterations, and the stopping
criterion is still an open problem. To avoid these two disadvantages,
many algorithms have been proposed~\cite{pasqualetti2012distributed,chen2002estimation,calafiore2009distributed,ajgl2014linear}.
One of the most important works is the one in~\cite{pasqualetti2012distributed},
where using the space structure of measurements and doing kernel projection,
each node achieves its minimum norm solution in a finite number of
steps.

In the static partially reconstructive problem, since each node considers
its own partial state, the consensus algorithm is not applicable.
For the case in which the graph induced by the communication network
is acyclic (i.e., without loops), an algorithm is proposed in~\cite{Taixin2013acc}.
In this algorithm, each node obtains a WLS estimate on its own state
in a finite number of steps. When the graph is cyclic (i.e., with
loops),~\cite{marelli2015distributed} gave a novel method which,
based on Richardson iterations, solves the WLS estimation problem.
However, it does so asymptotically, i.e., in infinite iterations.
We point out that most estimation algorithms for large-scale systems
are partially reconstructive, since the whole state of the system
is often of very high dimension.

In the dynamic fully reconstructive problem, the consensus algorithm
is also a popular option. In~\cite{matei2012}, one consensus algorithm
is run at time each sampling time, using the partial estimates obtained
at each node, based on their local measurements. Building on this
line, a study on the number of consensus iterations required at each
sampling time to guarantee the stability of the estimator, under the
observability condition, is done in~\cite{Acikmese2014}. Also, the
so-called diffusion Kalman filter~\cite{Cattivelli2010} runs consensus
on the estimates obtained at each sensor, using local measurements
as well as those from neighbors. As opposite to doing consensus on
the estimates, the authors of~\cite{Battistelli2014} found that,
by running consensus on the information matrices and vectors, observability
is sufficient for the estimation stability.

Concerning the dynamic partially reconstructive problem, information
passing and processing methods guaranteeing a stable estimate and
proposed in~\cite{zhou2013coordinated,zhou2015controllability,farina2010moving,khan2008distributing}.
Also, the authors of~\cite{haber2013moving} study systems with banded
dynamic state transition matrices, concluding that the contribution
from faraway nodes decreases with the increase of their distance.
The authors also propose the moving horizon estimation approach as
an approximation to the optimal state estimate.

In this paper we focus on the static partially reconstructive problem.
Also, as typically done in static problems, we assume that the vector
to be estimated is deterministic. More precisely, we consider the
algorithm in~\cite{Taixin2013acc}, which, as mentioned, yields the
optimal solution in finite-time, only when the communication graph
is acyclic. For cyclic graphs, this algorithm is not guaranteed to
produce the optimal solution. Nevertheless, in many applications,
even in the presence of loops, it delivers very good approximations
to the optimal solution, in only a very few steps. For those applications,
this makes the algorithm a valid alternative to the method in~\cite{marelli2015distributed}
even for cyclic networks. This is because, while the later guarantees
the optimal solution, the former one converges much faster. Motivated
by this, we study the accuracy of the estimate produced by the algorithm
in~\cite{Taixin2013acc}, under the general setting of a cyclic graph.

For a class of systems whose topological sparsity and signal-to-noise
ratio satisfy certain condition, we are able to determine the accuracy
of the estimates and their associated estimation error covariances,
with respect to those achievable via a centralized WLS method. Our
formulas clearly show how accuracy depends on the so-called loop-free
depth of each node. More precisely, the estimates and estimation error
covariances approach those from the centralized solution, exponentially
on the loop-free depth.

The rest of this paper is organized as follows. In Section~\ref{sec:formulation},
we give the problem formulation and introduce the distributed WLS
algorithm under study. In Section~\ref{sec:conver}, we show how
to convert a given graph into other equivalent ones, which are instrumental
for analyzing the behavior of the algorithm in cyclic graphs. In Section~\ref{sec:preliminary},
we introduce our notation, as well as the definition of the \emph{Riemannian
Distance} between matrices, together with some of its properties.
The accuracy of the information matrices (i.e., the inverses of the
error covariances) and state estimates produced by the distributed
WLS algorithm are analyzed in Sections~\ref{sec:Matrix} and~\ref{sec:Estimate},
respectively. In Section~\ref{sec:simulation}, we provide some simulations
to illustrate our results. Finally, concluding remarks are stated
in Section~\ref{sec:conclusion}. Complementary mathematical material
(including most proofs and some additional lemmas) appear in the Appendix.

\section{Problem Formulation\label{sec:formulation}}

Consider a system observed by $I$ sensing nodes. Associated to this
system, there is a deterministic vector $x^{T}=\left[x_{1}^{T},x_{2}^{T},\ldots,x_{I}^{T}\right]\in\mathbb{R}^{n}$,
with $\sum_{i=1}^{I}n_{i}=n$, called the global state. For any $i=1,\ldots,I$,
node~$i$ aims to estimate the sub-vector $x_{i}\in\mathbb{R}^{n_{i}}$.
There are also two kind of measurements. The so-called \emph{self
measurements} for node $i$ 
\begin{equation}
z_{i}=C_{i}x_{i}+v_{i},\label{eq:cycsys1}
\end{equation}
and the (pair-wise) \emph{joint measurements} between nodes $i$ and
$j$ 
\begin{equation}
z_{i,j}=C_{i,j}x_{i}+C_{j,i}x_{j}+v_{i,j}.\label{eq:cycsys2}
\end{equation}
In the above, the matrices $C_{i},C_{i,j}$ and $C_{j,i}$ are known,
and $v_{i}$ and $v_{i,j}$ are independent measurement noises with
known covariances $R_{i}>0$ and $R_{i,j}>0$, respectively. Note
that 1) the pair $(i,j)$ is unordered, i.e., $(i,j)=(j,i)$; 2) $z_{i,j}=z_{j,i}$
and $v_{i,j}=v_{j,i}$; 3) It is not necessary for all nodes to have
self measurements or all node pairs to have joint measurements. In
fact, joint measurements are typically sparse for large graphs.

We assume that node~$i$ and node~$j$ could communicate if $z_{i,j}$
exist. Furthermore, we call node~$j$ a neighbour of node~$i$ (i.e.,
$j\in\mathcal{N}_{i}$) and node~$i$ a neighbour of node~$j$ (i.e.,
$i\in\mathcal{N}_{j}$) if there is communication between them. In
view of this, communication between nodes is always two-ways; and
therefore, the associated communication graph (which will be formally
introduced later) is always undirected.

The target of distributed WLS estimation is to compute the WLS estimate
for each $x_{i}$, and its associated estimation error covariance,
using a fully distributed algorithm. The algorithm summarized in Algorithm~\ref{alg:BP},
achieves this goal. In this algorithm, at iteration $N$, node $i$
computes a local estimate $\hat{x}_{i}(N)$ of its sub-vector of interest,
and its associated covariance $\Sigma_{i}(N)$, using its local information
vector $\alpha_{i}(N)$ and information matrix $Q_{i}(N)$. Then,
for each neighbor $j\in\mathcal{N}_{i}$, it removes from $\alpha_{i}(N)$
and $Q_{i}(N)$ the information vector $\alpha_{j\rightarrow i}(N-1)$
and matrix $Q_{j\rightarrow i}(N-1)$, respectively, which it received
at the previous iteration from neighbor $j$, to built the information
vector $\alpha_{i\rightarrow j}(N)$ and matrix $Q_{i\rightarrow j}(N)$,
that it sends to the same neighbor at the current iteration.

\begin{algorithm}[t]
\caption{Distributed WLS algorithm.}
\label{alg:BP} \begin{algorithmic}

\STATE 1) \textbf{Initialization:} At time $k=0$, node $i$ defines:
\begin{align}
\alpha_{j\rightarrow i}(0) & =0,\quad Q_{j\rightarrow i}(0)=0.\label{newinitial}
\end{align}

\STATE 2) \textbf{Main loop:} At time $N=1,2,\cdots$, do:

2.1) Each node $i$ computes 
\begin{align}
\alpha_{i}(N) & =C_{i}^{T}R_{i}^{-1}z_{i}+\sum_{j\in\mathcal{N}_{i}}\alpha_{j\rightarrow i}(N-1),\nonumber \\
Q_{i}(N) & =C_{i}^{T}R_{i}^{-1}C_{i}+\sum_{j\in\mathcal{N}_{i}}Q_{j\rightarrow i}(N-1),\label{eq:BP-Q}
\end{align}
and 
\begin{align}
\hat{x}_{i}(N) & =Q_{i}^{-1}(N)\alpha_{i}(N),\quad\Sigma_{i}(N)=Q_{i}^{-1}(N).\label{eq:BP-x-hat}
\end{align}
2.2) Each node $i$ sends to each connected node $j$ with $j\in\mathcal{N}_{i}$:
\begin{align}
\alpha_{i\rightarrow j}(N) & =C_{j,i}^{T}R_{i\rightarrow j}^{-1}(N)z_{i\rightarrow j}(N),\nonumber \\
Q_{i\rightarrow j}(N) & =C_{j,i}^{T}R_{i\rightarrow j}^{-1}(N)C_{j,i},\label{eq:BP-Q2}
\end{align}
where 
\begin{align}
z_{i\rightarrow j}(N)= & z_{i,j}-C_{i,j}\left(Q_{i}(N)-Q_{j\rightarrow i}(N-1)\right)^{-1}\nonumber \\
 & \cdot(\alpha_{i}(N)-\alpha_{j\rightarrow i}(N-1)),\label{eq:BP-Q3a}\\
R_{i\rightarrow j}(N)= & R_{i,j}+C_{i,j}{(Q_{i}(N)-Q_{j\rightarrow i}(N-1))}^{-1}C_{i,j}^{T}.\label{eq:BP-Q3}
\end{align}

\end{algorithmic} 
\end{algorithm}

Algorithm~\ref{alg:BP} requires Assumption~\ref{asu:1}, which
is given below. This assumption implies that each node is able to
obtain a (possibly coarse) estimate of its sub-vector of interest,
using only its self measurements. Notice that, if this assumption
is not met, we have, at time $N=1$, and node $i$, that $Q_{i}(1)-Q_{j\rightarrow i}(0)=C_{i}^{T}R_{i}^{-1}C_{i}$.
Hence, $Q_{i}(1)-Q_{j\rightarrow i}(0)$ cannot be inverted in~\eqref{eq:BP-Q3a}
and~\eqref{eq:BP-Q3}.

Before stating Assumption~\ref{asu:1} we introduce some required
notation. 
\begin{notation}
The superscript $^{T}$ denotes vector or matrix transposition. For
a matrix $A$, $\|A\|$ denotes the induced operator norm, i.e., the
maximum singular value of $A$. Also, $A>0$ ($A\geq0$) means that
$A$ is positive definite (semi-definite), i.e., $x^{T}Ax>0$ ($x^{T}Ax\geq0$),
for all $x\neq0$. For a second matrix $B$, $A>B$ ($A\geq B$) means
that $A-B>0$ ($A-B\geq0$).
\end{notation}
\begin{assumption}
\label{asu:1} For every $i=1,2,\ldots,I$, we have 
\[
C_{i}^{T}R_{i}^{-1}C_{i}>0.
\]
\end{assumption}
It is known that Algorithm~\ref{alg:BP} converges to the correct
estimates in a finite number of iterations, when its associated communication
graph is acyclic~\cite{Taixin2013acc}. In fact, the required number
of iterations equals the diameter of the graph, i.e., the maximum
number of edges connecting one node to another over the whole graphs.
The fundamental challenge in our study is to understand how the algorithm
performs for cyclic graphs. As mentioned in Section~\ref{sec:Introduction},
the goal of this paper is to quantify the accuracy of the estimate
when the graph is cyclic, i.e., quantify the difference between the
distributed estimate and the centralized one. We split our accuracy
analysis in that of the information matrix (Section~\ref{sec:Matrix})
and that of the state estimate (Section~\ref{sec:Estimate}). In
the rest of paper, without loss of generality, we concentrate our
study on the accuracy of an arbitrary node, which is labeled as node~1.

\section{Graph Representations\label{sec:conver}}

In the network described above, nodes have only self and pair-wise
joint measurements. This permits using a simple connectivity (undirected)
graph, called \emph{canonical graph}, to describe the nodes and their
measurements. This is explained in Section~\ref{sub:Canonical-graph}.
A drawback of this representation for our intended analysis is that
this graph is cyclic in general. In Section~\ref{sub:conv-acyclic}
we describe how to convert this cyclic graph into an acyclic one,
with an infinite number of nodes. It turns out that this is an equivalent
graph, as far as distributed estimation is concerned. Then, in Section~\ref{sub:conv-line-2},
we explain how to further convert the acyclic graph into another equivalent
one, whose topology is that of a single line. For a more detailed
presentation of equivalent graph transformations, the reader is referred
to~\cite{Weiss2000graph,tatikonda2002loopy,tatikonda2003convergence}.
We point out that all the above graphs are undirected.

\subsection{Canonical graph representation\label{sub:Canonical-graph}}

The canonical graph $\mathcal{G}$ has a node associated with each
sensing node $i=1,\ldots,I$. Also, nodes $i$ and $j$ are connected
by an undirected edge if they can communicate with each other, i.e.,
$j\in\mathcal{N}_{i}$.

\subsection{Acyclic graph representation\label{sub:conv-acyclic}}

We start the section with the following definition.
\begin{defn}
A rooted tree graph is an acyclic connected graph, in which a node
is assigned as its root. For a node $i$ different from the root one,
we let $p(i)$ denote its \emph{parent} (i.e., the next node when
moving towards the root) and $\mathcal{S}_{i}$ denote the set of
its \emph{children} (i.e., all the nodes $j$ with $i=p(j)$). Also,
node $i$ is called a leaf if $\mathcal{S}_{i}$ is empty.
\end{defn}
Given a cyclic canonical graph $\mathcal{G}$, we can convert it into
an acyclic one $\mathcal{A}$, having a rooted tree topology, with
any arbitrary node as its root one. Since, as pointed out before,
we concentrate our study on the accuracy at node~1, we choose this
node as the root one. This graph enjoys the property that, if Algorithm~\ref{alg:BP}
is applied to both graphs, it will produce at node $1$ and iteration
$N$, the same result.

Graph $\mathcal{A}$ has an infinite number of nodes. Each of its
nodes is associated to a node in $\mathcal{G}$. With some abuse of
notation, we use $\mathcal{G}(n)$ to denote the node in $\mathcal{G}$
associated to node $n$ in $\mathcal{A}$, and $\mathcal{A}(i)$ to
denote the set of nodes in $\mathcal{A}$ associated to node $i$
in $\mathcal{G}$. Graph $\mathcal{A}$ is constructed as the limit
of the following iterative procedure. We start by defining $\mathcal{A}_{0}$
as the empty graph and $\mathcal{A}_{1}$ as the graph having no edges,
and having a single node, which is associated to node~$1$ of $\mathcal{G}$.
Then, at each step $N\geq2$, we do the following steps:
\begin{enumerate}
\item Find all leaf nodes $l$ of the tree $\mathcal{A}_{N-1}$.
\item Find all neighbors $j$ of $\mathcal{G}(l)$ in $\mathcal{G}$, excluding
all nodes in $\mathcal{A}(p(l))$ (i.e., associated to the parent
of $l$ in $\mathcal{A}$).
\item For each $l$ and $j$:
\begin{enumerate}
\item Add a node $n$ to the tree.
\item Add the undirected edge from $l$ to $n$ to the tree.
\item Associate $n$ in $\mathcal{A}$ to $j$ in $\mathcal{G}$. 
\end{enumerate}
\item Define $\mathcal{A}_{N}$ as the resulting graph.
\end{enumerate}
Our next step is to associate a system of measurement equations to
the acyclic graph $\mathcal{A}$, in a way similar to the way in which
the system of equations~(\ref{eq:cycsys1})-(\ref{eq:cycsys2}) is
associated with $\mathcal{G}$. These equations need to satisfy two
conditions. First, their canonical graph should be $\mathcal{A}$.
Second, the aforementioned equivalence at node $1$ should be preserved.
As explained in~\cite{ihler2005loopy,Weiss2000graph,tatikonda2002loopy,tatikonda2003convergence},
both conditions are satisfied if 
\begin{eqnarray}
\bar{z}_{i} & = & \bar{C}_{i}\bar{x}_{i}+\bar{v}_{i},\label{eq:acycsys1}\\
\bar{z}_{i,j} & = & \bar{C}_{i,j}\bar{x}_{i}+\bar{C}_{j,i}\bar{x}_{j}+\bar{v}_{i,j}\label{eq:acycsys2}
\end{eqnarray}
for all $i\in\mathcal{A}$ and $j\in\mathcal{S}_{i}$, with $\bar{v}_{i}\sim\mathcal{N}\left(0,\bar{R}_{i}\right)$
and $\bar{v}_{i,j}\sim\mathcal{N}\left(0,\bar{R}_{i,j}\right)$. The
values of the quantities in~\eqref{eq:acycsys1}-\eqref{eq:acycsys2}
are given by those corresponding to the nodes in $\mathcal{G}$ which
are associated to nodes $i$ and $j$ in $\mathcal{A}$, i.e.,
\begin{align*}
\bar{z}_{i} & =z_{\mathcal{G}(i)},\quad\bar{v}_{i}=v_{\mathcal{G}(i)},\quad\bar{C}_{i}=C_{\mathcal{G}(i)},\\
\bar{R}_{i} & =R_{\mathcal{G}(i)},\quad\bar{z}_{i,j}=z_{\mathcal{G}(i),\mathcal{G}(j)},\quad\bar{v}_{i,j}=v_{\mathcal{G}(i),\mathcal{G}(j)},\\
\bar{C}_{i,j} & =C_{\mathcal{G}(i),\mathcal{G}(j)},\quad\bar{R}_{i,j}=R_{\mathcal{G}(i),\mathcal{G}(j)}.
\end{align*}
Also, all noises $\bar{v}_{i}$ and $\bar{v}_{i,j}$, $i=1,\cdots,I$,
$j\in\mathcal{N}_{i}$, are pairwise uncorrelated.
\begin{rem}
\label{rem:WLS-eq}Recall that, in an acyclic graph, Algorithm~\ref{alg:BP}
produces, at each node, the same estimate that would be obtained using
the centralized WLS method. Since the graph $\mathcal{A}_{N}$ is
acyclic, the outcomes of both methods will be the same on $\mathcal{A}_{N}$.
Moreover, its measurements~(\ref{eq:acycsys1})-(\ref{eq:acycsys2}),
are designed so that, at node~$1$ and iteration $N$, this outcome
equals that resulting from applying Algorithm~\ref{alg:BP} to graph
$\mathcal{G}$. Hence, $\mathcal{G}$ and $\mathcal{A}_{N}$ are equivalent
graphs only from the point of view Algorithm~\ref{alg:BP} (at node~$1$
and iteration $N$), but not from that of centralized WLS. 
\end{rem}

\subsection{Representation as a line graph\label{sub:conv-line-2}}

Let $\mathcal{A}_{N}$ be the $N$-layer acyclic graph with root node
1 and measurement equations~(\ref{eq:acycsys1})-(\ref{eq:acycsys2}),
as described above. We now describe how to convert $\mathcal{A}_{N}$
into a line graph $\mathcal{L}_{N}$ such that the aforementioned
equivalence is still preserved.

Indeed, $\mathcal{L}_{N}$ is formed by simply grouping all the nodes
in $\mathcal{A}_{N}$, which are exactly $n-1$ hops away from node
$1$, into a super node $\mathcal{T}_{n}$, for all $n=1,2,\ldots,N$.
In particular, $\mathcal{T}_{1}$ is just node 1. 

Again, we need to associate a system of measurement equations to $\mathcal{L}_{N}$
satisfying the conditions described in Section~\ref{sub:conv-acyclic}.
This is done by grouping all the measurement equations for each super
node $\mathcal{T}_{n}$, as detailed below.

Denote the size of any finite set $S$ by $|S|$ and its elements
by $S(1),S(2),\ldots,S(|S|)$. For each $n\in\mathbb{N}$, the state
of $\mathcal{T}_{n}$ is given by 
\begin{eqnarray*}
\tilde{x}_{n} & = & \begin{bmatrix}\bar{x}_{\mathcal{T}_{n}(1)}^{T}, & \bar{x}_{\mathcal{T}_{n}(2)}^{T}, & \ldots, & \bar{x}_{\mathcal{T}_{n}(|\mathcal{T}_{n}|)}^{T}\end{bmatrix}^{T},
\end{eqnarray*}
and its measurement equations are given by 
\begin{eqnarray}
\tilde{z}_{n} & = & \tilde{C}_{n}\tilde{x}_{n}+\tilde{v}_{n},\label{eq:linesys3}\\
\tilde{z}_{n,n+1} & = & \tilde{C}_{n,n+1}\tilde{x}_{n}+\tilde{C}_{n+1,n}\tilde{x}_{n+1}+\tilde{v}_{n,n+1}.\label{eq:linesys4}
\end{eqnarray}
That is, $\tilde{z}_{n}$ consists of all the measurements $\bar{z}_{i}$
with $i\in\mathcal{T}_{n}$, and $\tilde{z}_{n,n+1}$ consists of
all the measurements $\bar{z}_{i,j}$ with $i\in\mathcal{T}_{n}$
and $j\in\mathcal{T}_{n+1}$. Note that $\tilde{v}_{n}\sim\mathcal{N}(0,\tilde{R}_{n})$
and $\tilde{v}_{n,n+1}\sim\mathcal{N}(0,\tilde{R}_{n,n+1})$. The
matrices $\tilde{C}_{n}$, $\tilde{R}_{n}$, $\tilde{C}_{n,n+1}$
and $\tilde{R}_{n,n+1}$ are naturally related to $\bar{C}_{i}$,
$\bar{R}_{i}$, $\bar{C}_{i,j}$ and $\bar{R}_{i,j}$ through the
above construction. More precisely, 
\begin{eqnarray*}
\tilde{z}_{n} & = & \begin{bmatrix}\bar{z}_{\mathcal{T}_{n}(1)}^{T}, & \bar{z}_{\mathcal{T}_{n}(2)}^{T}, & \ldots, & \bar{z}_{\mathcal{T}_{n}(|\mathcal{T}_{n}|)}^{T}\end{bmatrix}^{T},\\
\tilde{C}_{n} & = & \mathrm{diag}\{\bar{C}_{\mathcal{T}_{n}(1)},\bar{C}_{\mathcal{T}_{n}(2)},\ldots,\bar{C}_{\mathcal{T}_{n}(|\mathcal{T}_{n}|)}\},\\
\tilde{R}_{n} & = & \mathrm{diag}\{\bar{R}_{\mathcal{T}_{n}(1)},\bar{R}_{\mathcal{T}_{n}(2)},\ldots,\bar{R}_{\mathcal{T}_{n}(|\mathcal{T}_{n}|)}\}.
\end{eqnarray*}
Similarly, 
\begin{eqnarray*}
\tilde{z}_{n,n+1} & = & \begin{bmatrix}\grave{z}_{\mathcal{T}_{n}(1)}^{T}, & \grave{z}_{\mathcal{T}_{n}(2)}^{T}, & \ldots, & \grave{z}_{\mathcal{T}_{n}(|\mathcal{T}_{n}|)}^{T}\end{bmatrix}^{T},\\
\tilde{C}_{n,n+1} & = & \mathrm{diag}\{\grave{C}_{\mathcal{T}_{n}(1)},\grave{C}_{\mathcal{T}_{n}(2)},\ldots,\grave{C}_{\mathcal{T}_{n}(|\mathcal{T}_{n}|)}\},\\
\tilde{C}_{n+1,n} & = & \mathrm{diag}\{\acute{C}_{\mathcal{T}_{n+1}(1)},\acute{C}_{\mathcal{T}_{n+1}(2)},\ldots,\acute{C}_{\mathcal{T}_{n+1}(|\mathcal{T}_{n+1}|)}\},\\
\tilde{R}_{n,n+1} & = & \mathrm{diag}\{\grave{R}_{\mathcal{T}_{n}(1)},\grave{R}_{\mathcal{T}_{n}(2)},\ldots,\grave{R}_{\mathcal{T}_{n}(|\mathcal{T}_{n}|)}\},
\end{eqnarray*}
with 
\begin{eqnarray*}
\grave{z}_{i} & = & \begin{bmatrix}\bar{z}_{i,\mathcal{S}_{i}(1)}^{T}, & \bar{z}_{i,\mathcal{S}_{i}(2)}^{T}, & \ldots, & \bar{z}_{i,\mathcal{S}_{i}(|\mathcal{S}_{i}|)}^{T}\end{bmatrix}^{T},\\
\grave{C}_{i} & = & \begin{bmatrix}\bar{C}_{i,\mathcal{S}_{i}(1)}^{T}, & \bar{C}_{i,\mathcal{S}_{i}(2)}^{T}, & \ldots, & \bar{C}_{i,\mathcal{S}_{i}(|\mathcal{S}_{i}|)}^{T}\end{bmatrix}^{T},\\
\acute{C}_{i} & = & \mathrm{diag}\{\bar{C}_{\mathcal{S}_{i}(1),i},\bar{C}_{\mathcal{S}_{i}(2),i},\ldots,\bar{C}_{\mathcal{S}_{i}(|\mathcal{S}_{i}|),i}\},\\
\grave{R}_{i} & = & \mathrm{diag}\{\bar{R}_{i,\mathcal{S}_{i}(1)},\bar{R}_{i,\mathcal{S}_{i}(2)},\ldots,\bar{R}_{i,\mathcal{S}_{i}(|\mathcal{S}_{i}|)}\}.
\end{eqnarray*}

\begin{rem}
Note that the statement in Remark~\ref{rem:WLS-eq} also holds for
$\mathcal{L}_{N}$. More precisely, the centralized WLS estimate at
node 1 in $\mathcal{L}_{N}$ equals to that of Algorithm~\ref{alg:BP},
when applied to $\mathcal{G}$, at the same nodes and iteration $N$. 
\end{rem}
In view of the above analysis, the problem of studying the dynamics
of Algorithm~\ref{alg:BP} becomes the problem of studying the centralized
WLS estimate at node~$1$ for the graph $\mathcal{L}_{N}$, as $N\rightarrow\infty$.

\section{Preliminaries\label{sec:preliminary}}

In our analysis below, we will make use of the so-called \emph{Riemannian
distance} between matrices~\cite{bougerol1993kalman}.
\begin{defn}
\label{def:rie} For $n\times n$ matrices $P,Q>0$, their Riemannian
distance is defined by 
\[
\delta\left(P,Q\right)=\sqrt{\sum_{k=1}^{n}\log^{2}\sigma_{k}\left(PQ^{-1}\right)}\ ,
\]
where $\sigma_{1}\left(X\right)\geq\cdots\geq\sigma_{n}\left(X\right)$
denote the singular values of matrix $X$. 
\end{defn}
The following proposition states a number of properties of the Riemannian
distance. Its proof appears in the appendix.
\begin{prop}
\label{prop:riemannian} For any $n\times n$ positive definite matrices
$P$ and $Q$, the following results hold: 
\end{prop}
\begin{enumerate}
\item $\delta(P,P)=0$. 
\item $\delta\left(P^{-1},Q^{-1}\right)=\delta\left(Q,P\right)=\delta\left(P,Q\right).$ 
\item If $B$ has full row rank, $\delta\left(BPB^{T},BQB^{T}\right)\leq\delta\left(P,Q\right)$,
and the equality holds if $B$ is invertible. 
\item If $P\ge Q$ and $W\geq0$, then $\delta\left(P+W,Q\right)\geq\delta\left(P,Q\right).$ 
\item For any $m\times m$ matrix $W>0$ and $m\times n$ matrix $B$, we
have 
\[
\delta(W+BP^{-1}B^{T},W+BQ^{-1}B^{T})\leq\frac{\alpha}{\alpha+\beta}\delta(P,Q),
\]
where $\alpha=\max\{\|BP^{-1}B^{T}\|,\|BQ^{-1}B^{T}\|\}$ and $\beta=\sigma_{\min}\left(W\right)$,
with $\sigma_{\min}(W)$ denoting the smallest singular value of $W$. 
\item If $P>Q$, then $\left\Vert P-Q\right\Vert \leq\left(e^{\delta\left(P,Q\right)}-1\right)\left\Vert Q\right\Vert .$ 
\end{enumerate}
We now introduce some notation that will be used in the rest of the
paper.
\begin{notation}
For a graph $\mathcal{C}$, we use $\hat{x}_{i}(\mathcal{C})$, $Q_{i}(\mathcal{C})$
and $\alpha_{i}(\mathcal{C})$ to denote the final (after convergence)
state estimate, information matrix and information vector, respectively,
obtained by running Algorithm~\ref{alg:BP} on $\mathcal{C}$. 
\end{notation}
\begin{notation}
\label{initial} Let $Q_{(i,j),0}\geq0$, for each $i=1,\cdots,I$,
$j\in\mathcal{N}_{i}$, and define the set $\mathbb{Q}=\left\{ Q_{(i,j),0}:i=1,2,\ldots,I,j\in\mathcal{N}_{i}\right\} $.
Suppose that we run Algorithm~\ref{alg:BP} on the network $\mathcal{G}$,
but replacing the initialization~\eqref{newinitial} by $Q_{j\rightarrow i}(0)=Q_{(i,j),0}$,
for all $i=1,\cdots,I$, $j\in\mathcal{N}_{i}$. We use $\hat{x}_{1}(N,\mathbb{Q})$
and $Q_{1}(N,\mathbb{Q})$ to denote the estimate and information
matrix, respectively, yield by such algorithm at node~$1$ and step
$N$. 
\end{notation}
\begin{notation}
\label{l_1} If $z_{i,j}$ exists, the pair $(i,j)$ is called an
(undirected) \emph{edge.} Notice that since edges are undirected,
the pairs $(i,j)$ and $(j,i)$ denote the same edge. A \emph{path}
is a concatenation of contiguous edges, and its \emph{length} is the
number of edges forming it. A \emph{cycle} is a path with no repetitions
of vertices and edges, except for the necessary repetition of the
starting and ending vertices. For each $i,j\in\{1,\cdots,I\}$, the
\emph{distance} between nodes~$i$ and~$j$ is defined as the minimum
length of a path joining these two nodes. Let $\mathcal{N}_{1}(l)$
denote the subgraph of $\mathcal{G}$ formed by nodes whose distance
from node~$1$ is less than or equal to $l$. The loop-free depth
$l_{1}$ of node~1 is the largest integer such that $\mathcal{N}_{1}(l_{1})$
is acyclic (i.e., without \emph{cycle}s). 
\end{notation}
We also introduce the following constants 
\begin{eqnarray*}
\bar{u} & := & \max_{i}|\mathcal{N}_{i}|-1,\quad\bar{n}:=\max_{i}\dim x_{i},\\
\bar{m} & := & \max\{\max_{i}\dim z_{i},\max_{i,j}\dim z_{i,j}\}.
\end{eqnarray*}

\section{Accuracy Analysis for the Information Matrix \label{sec:Matrix}}

In this section we derive a bound for the difference between the information
matrix yield by the Algorithm~\ref{alg:BP} and that obtained using
the centralized WLS method. For a class of system, we provide a lower
bound for the time after which the difference falls within this bound.
Moreover, this bound decreases exponentially with the increase of
the loop free-depth. Our main result is given in the Subsection~\ref{subsec:main1}.
Its proof appears in Subsection~\ref{subsec:proof}.

\subsection{Main Result\label{subsec:main1}}

The main result on the accuracy of the information matrix is given
below.
\begin{thm}
\label{disLSE} Let $\mathrm{Cov}_{1}^{\mathrm{WLS}}$ be the estimation
error covariance obtained at node~1 when using centralized WLS. If
$\rho<1$, then there exists a constant $\varpi$ (only dependent
on the system parameters $\bar{u}$, $\bar{n}$, $C_{i,j}$, $C_{i}$,
$R_{i,j}$ and $R_{i}$) such that, for any $N\geq\textit{l}_{1}+1$,
\[
\Vert\mathrm{Cov}_{1}^{\mathrm{WLS}}-Q_{1}^{-1}(N)\Vert\leq\varpi\rho^{\textit{l}_{1}},
\]
where 
\begin{align*}
\rho & =\lambda\sqrt{\bar{u}},\quad\lambda=\frac{\alpha_{1}}{\alpha_{1}+\beta_{1}}\frac{\alpha_{2}}{\alpha_{2}+\beta_{2}},\\
\alpha_{1} & =\bar{u}\max_{i,j}\Vert C_{i,j}^{T}R_{i,j}^{-1}C_{i,j}\Vert,\quad\beta_{1}=\min_{i}\sigma_{\min}(C_{i}^{T}R_{i}^{-1}C_{i}),\\
\alpha_{2} & =\max_{i,j}\Vert C_{i,j}(C_{i}^{T}R_{i}^{-1}C_{i})^{-1}C_{i,j}^{T}\Vert,\beta_{2}=\min_{i,j}\sigma_{\min}(R_{i,j}).
\end{align*}
\end{thm}
\begin{rem}
Theorem~\ref{disLSE} states that, if the graph/system satisfies
$\rho<1$, the inverse of the information matrix $Q_{1}^{-1}(N)$
yield by Algorithm~\ref{alg:BP} at node~1, exponentially approaches
the estimation error covariance of centralized WLS, as its loop-free
depth $\textit{l}_{1}$ increases. 
\end{rem}
\begin{rem}
\label{rem:system-class} Since $\rho=\lambda\sqrt{\bar{u}}$, the
result is mainly given for a class of graph with sparse connections
(small $\bar{u}$) and where the ratio between the signal-to-noise-ratio
(SNR) of local measurements and the SNR of each joint measurement
is small (small $\lambda$). Notice that the later condition is relatively
mild, since each joint measurement typically has a lower SNR than
the local one. 
\end{rem}

\subsection{Proof of Theorem~\ref{disLSE} \label{subsec:proof}}

Recall that, in view of the graph conversions described in Section~\ref{sec:conver},
we have $Q_{1}(N)=Q_{1}(\mathcal{A}_{N})=Q_{1}(\mathcal{L}_{N})$.
The proof of Theorem~\ref{disLSE} uses this fact. It also requires
the following lemmas, whose proofs appear in the Appendix.
\begin{lem}
\label{lem:decay-Q-1} Let $\mathbb{Q}_{1}$ and $\mathbb{Q}_{2}$
be initial sets both satisfying $0\leq Q_{(i,j),0}^{c}\leq C_{i,j}^{T}R_{i,j}^{-1}C_{i,j}$,
for all $Q_{(i,j),0}^{c}\in\mathbb{Q}_{c}$ and $c\in\left\{ 1,2\right\} $.
Then, in the notation of Theorem~\ref{disLSE}, for all $N\in\mathbb{N}$,
\[
\delta\left(Q_{1}(N,\mathbb{Q}_{1})-Q_{1}(N,\mathcal{\mathbb{Q}}_{2})\right)\leq\rho^{N-1}\bar{\delta},
\]
with 
\begin{multline*}
\bar{\delta}=\sqrt{(\bar{u}+1)\bar{n}}\\
\times\max_{i}\log\Vert I+(\sum_{j\in\mathcal{N}_{i}}C_{i,j}^{T}R_{i,j}^{-1}C_{i,j})(C_{i}^{T}R_{i}^{-1}C_{i})^{-1}\Vert.
\end{multline*}
\end{lem}
\begin{notation}
Let 
\[
\mathbb{Q}^{\mathrm{M}}=\left\{ Q_{(i,j),0}^{\mathrm{M}}:i=1,\ldots,I~\text{and}~j\in\mathcal{N}_{i}\right\} ,
\]
\[
\mathbb{Q}^{0}=\left\{ Q_{(i,j),0}^{0}:i=1,\ldots,I~\text{and}~j\in\mathcal{N}_{i}\right\} ,
\]
with 
\begin{align*}
Q_{(i,j),0}^{\mathrm{M}} & =C_{i,j}^{T}R_{i,j}^{-1}C_{i,j},\quad Q_{(i,j),0}^{0}=0.
\end{align*}
In particular, notice that $\mathbb{Q}^{0}$ is the initialization
used in Algorithm~\ref{alg:BP}, i.e., $\hat{x}_{1}(N)=\hat{x}_{1}(N,\mathbb{Q}^{0})$
and $Q_{1}(N)=Q_{1}(N,\mathbb{Q}^{0})$. 
\end{notation}
\begin{lem}
\label{lem:accu-Q} Recall the definition of loop-free depth $\textit{l}_{1}$
from Notation~\ref{l_1}. For any $N\geq\textit{l}_{1}+1$, we have
\begin{multline}
\left\Vert Q_{1}^{-1}(N)-\mathrm{Cov}_{1}^{\mathrm{WLS}}\right\Vert \\
\leq\left\Vert Q_{1}^{-1}(\textit{l}_{1}+1,\mathbb{Q}^{\mathrm{M}})-Q_{1}^{-1}(\textit{l}_{1}+1,\mathbb{Q}^{0})\right\Vert .\label{ineq}
\end{multline}
\end{lem}
The proof of Theorem~\ref{disLSE} uses the above property to provide
an upper bound for the difference between $Q_{1}^{-1}(N)$ and $\mathrm{Cov}_{1}^{\mathrm{WLS}}$.
\begin{pf}
(of Theorem~\ref{disLSE}) Since both $\mathbb{Q}^{\mathrm{M}}$
and $\mathbb{Q}^{0}$ satisfy the condition in Lemma~\ref{lem:decay-Q-1},
it follows that 
\begin{align*}
 & \delta(Q_{1}^{-1}(\textit{l}_{1}+1,\mathbb{Q}^{0}),Q_{1}^{-1}(\textit{l}_{1}+1,\mathbb{Q}^{\mathrm{M}}))\leq\rho^{\textit{l}_{1}}\bar{\delta}.
\end{align*}
From the Proposition~\ref{prop:riemannian}, 
\begin{multline*}
\Vert Q_{1}^{-1}(\textit{l}_{1}+1,\mathbb{Q}^{0}),Q_{1}^{-1}(\textit{l}_{1}+1,\mathbb{Q}^{\mathrm{M}})\Vert\\
\leq(e^{\rho^{\textit{l}_{1}}\bar{\delta}}-1)\Vert Q_{1}^{-1}(\textit{l}_{1}+1)\Vert\leq(e^{\rho^{\textit{l}_{1}}\bar{\delta}}-1)\Vert(C_{1}^{T}R_{1}^{-1}C_{1})^{-1}\Vert\\
\leq\beta_{1}^{-1}(e^{\bar{\delta}}-1)\rho^{\textit{l}_{1}},
\end{multline*}
where the last inequality follows from Lemma~\ref{lem:exp-bound}.
Since the quantity $\beta_{1}^{-1}(e^{\bar{\delta}}-1)$ depends only
on $\bar{u},\bar{n}$ and the system parameters $C_{i,j},C_{i},R_{i,j},R_{i}$
for some $i$ and $j$, the result then follows from~\eqref{ineq}.
\end{pf}

\section{Accuracy Analysis for the State Estimate \label{sec:Estimate}}

In this section, we derive a bound for the difference between the
estimate yield by Algorithm~\ref{alg:BP} and that obtained using
the centralized WLS method. For a class of system, we provide a lower
bound for the time after which the difference falls within this bound.
Moreover, this bound decreases exponentially with the increase of
the loop-free depth. Our main result is given in the Subsection~\ref{subsec:estimate}
and its proof appears in Subsection~\ref{subsec:proof2}.

\subsection{Main Result\label{subsec:estimate}}

The main result on the accuracy of the estimate is given below.
\begin{thm}
\label{xaccuracy} Let $\hat{x}_{1}^{\mathrm{WLS}}$ be the estimate
obtained at node~1 when using centralized WLS. If ${\kappa}<1$,
then there exists a constant $\varpi$ (only dependent on the system
parameters $\bar{m}$, $\bar{u}$, $\bar{n}$, $C_{i,j}$, $C_{i}$,
$R_{i,j}$ and $R_{i}$ as well as on the measurements $z_{i,j}$,
$z_{i}$) such that, for all $N\geq\textit{l}_{1}+1$, 
\[
\Vert\hat{x}_{1}(N)-\hat{x}_{1}^{\mathrm{WLS}}\Vert\leq\varpi\kappa^{\textit{l}_{1}+1}
\]
where with 
\begin{align*}
\kappa & =\max\{\bar{u}\sqrt{\omega},\sqrt{\bar{u}}\iota^{1/\zeta}\},\quad\omega=\frac{a_{1}}{a_{1}+b_{1}}\frac{a_{2}}{a_{2}+b_{2}},\\
a_{1} & =\underline{r}^{-1}\bar{u}\max_{i,j}\left\Vert C_{i,j}\right\Vert ^{2},\quad a_{2}=\max_{i,j}\Vert C_{i,j}\Vert^{2}\frac{\bar{u}\bar{r}}{\underline{\varepsilon}^{2}},\\
b_{1} & =\bar{r}^{-1}\underline{\varepsilon}^{2},\quad b_{2}=\underline{r},\quad\iota=\frac{\sqrt{\overline{q}}-\sqrt{\underline{q}}}{\sqrt{\overline{q}}+\sqrt{\underline{q}}},\\
\overline{q} & =\overline{\varepsilon}^{2}\underline{r}^{-1},\quad\underline{q}=\underline{\varepsilon}^{2}\overline{r}^{-1},\quad\zeta=2+\log_{\frac{1}{\sqrt{\omega}}}(\overline{q}/\underline{q}).\\
\overline{r} & =\max_{i}\{\|R_{i}\|,\|R_{i,j}\|\},\underline{r}=\min_{i}\{\sigma_{\min}(R_{i}),\sigma_{\min}(R_{i,j})\},\\
\overline{\varepsilon} & =\max_{i,j}\sqrt{\left\Vert C_{i}\right\Vert ^{2}+4\bar{u}\left\Vert C_{i,j}\right\Vert ^{2}},\quad\underline{\varepsilon}=\min_{i}\sigma_{\min}(C_{i}).
\end{align*}
\end{thm}
\begin{rem}
From the definition of ${\kappa}$, the observation made in Remark~\ref{rem:system-class}
also applies to Theorem~\ref{xaccuracy}. However, since ${\kappa}\geq\rho$,
the condition required for Theorem~\ref{xaccuracy} is stronger than
that for Theorem~\ref{disLSE}. 
\end{rem}

\subsection{Proof of Theorem~\ref{xaccuracy} \label{subsec:proof2}}

We split the proof of Theorem~\ref{xaccuracy} into three parts.
In Section~\ref{sub:est-line} we derive a bound for the increment
$\hat{x}_{1}(N+1)-\hat{x}_{1}(N)$ in a graph with line topology.
In Section~\ref{sub:est-arbitrary} we generalize this result for
an arbitrary graph. Finally, in Section~\ref{sub:est-acc} we use
this result to bound the difference between the state estimate yield
by the Algorithm~\ref{alg:BP} and that obtained using the centralized
WLS method.

\subsubsection{Bound of the increment in a line graph \label{sub:est-line}}

Consider a line graph $\mathcal{L}_{N}$ with measurement equations
given by~\eqref{eq:linesys3}-\eqref{eq:linesys4}. Let 
\begin{eqnarray}
y_{i} & = & \left[\tilde{z}_{i}^{T},\tilde{z}_{i,i+1}^{T}\right]^{T},\quad w_{i}=\left[\tilde{v}_{i}^{T},\tilde{v}_{i,i+1}^{T}\right]^{T},\nonumber \\
A_{i,i} & = & \begin{bmatrix}\tilde{C}_{i}\\
\tilde{C}_{i,i+1}
\end{bmatrix},\quad A_{i,i+1}=\begin{bmatrix}0\\
\tilde{C}_{i+1,i}
\end{bmatrix},\nonumber \\
S_{i} & = & \begin{bmatrix}\tilde{R}_{i} & 0\\
0 & \tilde{R}_{i,i+1}
\end{bmatrix}\label{eq:S}
\end{eqnarray}
for any $i=1,2,\ldots,N-1$, and 
\[
y_{N}=\tilde{z}_{N},\quad w_{i}=\tilde{v}_{N},\quad A_{N,N}=\tilde{C}_{N},\quad S_{N}=\tilde{R}_{N}.
\]
Then,~\eqref{eq:linesys3} and~\eqref{eq:linesys4} become 
\[
y_{i}=A_{i,i}\tilde{x}_{i}+A_{i,i+1}\tilde{x}_{i+1}+w_{i},
\]
with $w_{i}\sim\mathcal{N}\left(0,S_{i}\right)$. We also define 
\begin{eqnarray*}
\mathbf{x}_{N} & = & \left[\tilde{x}_{1}^{T},\cdots,\tilde{x}_{N}^{T}\right]^{T},\ \mathbf{y}_{N}=\left[y_{1}^{T},\cdots,y_{N}^{T}\right]^{T},\\
\mathbf{w}_{N} & = & \left[w_{1}^{T},\cdots,w_{N}^{T}\right]^{T},\\
\left[\mathbf{A}_{N}\right]_{n,m} & = & \begin{cases}
A_{n,m}, & 0\leq m-n\leq1,\\
0, & \text{otherwise.}
\end{cases}
\end{eqnarray*}
We then have 
\[
\mathbf{y}_{N}=\mathbf{A}_{N}\mathbf{x}_{N}+\mathbf{w}_{N},
\]
with $\mathbf{w}_{N}\sim\mathcal{N}\left(0,\mathbf{S}_{N}\right)$
and $\mathbf{S}_{N}=\mathrm{diag}\left\{ S_{1},\cdots,S_{N}\right\} $.

The WLS estimate $\hat{\mathbf{x}}_{N}$ of $\mathbf{x}_{N}$ is given
by 
\[
\hat{\mathbf{x}}_{N}=\mathbf{Q}_{N}^{-1}\mathbf{q}_{N},
\]
where $\mathbf{q}_{N}=\mathbf{A}_{N}^{T}\mathbf{S}_{N}^{-1}\mathbf{y}_{N}=\left[q_{1}^{T},\cdots,q_{N}^{T}\right]^{T}$
with 
\begin{equation}
q_{i}=\begin{cases}
A_{i,i}^{T}S_{i}^{-1}y_{i}, & i=1,\\
A_{i,i}^{T}S_{i}^{-1}y_{i}+A_{i-1,i}^{T}S_{i-1}^{-1}y_{i-1}, & i>1,
\end{cases}\label{eq:q_n}
\end{equation}
and the $(i,j)$-th entry $Q_{i,j}$ of $\mathbf{Q}_{N}=\mathbf{A}_{N}^{T}\mathbf{S}_{N}^{-1}\mathbf{A}_{N}$
given by 
\begin{align}
Q_{i,i} & =\begin{cases}
A_{i,i}^{T}S_{i}^{-1}A_{i,i}, & i=1,\\
A_{i,i}^{T}S_{i}^{-1}A_{i,i}+A_{i-1,i}^{T}S_{i-1}^{-1}A_{i-1,i}, & i>1,
\end{cases}\nonumber \\
Q_{i,i+1} & =A_{i,i}^{T}S_{i}^{-1}A_{i,i+1},~Q_{i+1,i}=Q_{i,i+1}^{T},\label{eq:Q-2}\\
Q_{i,j} & =0,\,\left|i-j\right|\geq2.\nonumber 
\end{align}
Let $\Sigma_{N}=\mathbf{Q}_{N}^{-1}$ and $[\Sigma_{N}]_{i,j}$ be
its $(i,j)$-th block. From the inverse formula for band matrices
given in Theorem~3.1 of~\cite{meurant1992review}, it follows that
the first block row of $\Sigma_{N}$ is given by 
\begin{equation}
\left[\Sigma_{N}\right]_{1,j}=\left(\prod_{k=1}^{j-1}\Delta_{k}^{-1}Q_{k,k+1}\right)\Phi_{j}^{-1}(N)\label{guanjianyinsu}
\end{equation}
with 
\begin{eqnarray}
\Phi_{j}(N) & = & \Gamma_{j}(N)-Q_{j,j-1}\Delta_{j-1}^{-1}Q_{j-1,j},\label{eq:Phi}\\
\Delta_{k} & = & \begin{cases}
Q_{kk}, & k=1,\\
Q_{kk}-Q_{k,k-1}\Delta_{k-1}^{-1}Q_{k-1,k}, & k>1,
\end{cases}\nonumber \\
\Gamma_{k}(N) & = & \begin{cases}
Q_{kk}, & k=N,\\
Q_{kk}-Q_{k,k+1}\Gamma_{k+1}^{-1}(N)Q_{k+1,k}, & k<N,
\end{cases}\nonumber 
\end{eqnarray}
for any $j=1,2,\ldots,N$. Then, the first entry $\left[\hat{\mathbf{x}}_{N}\right]_{1}$
of $\hat{\mathbf{x}}_{N}$ is given by 
\begin{equation}
\left[\hat{\mathbf{x}}_{N}\right]_{1}=\sum_{j=1}^{N}\left[\Sigma_{N}\right]_{1,j}q_{j}.\label{eq:x1-hat-WLS}
\end{equation}
Recall that $\hat{x}_{1}(\mathcal{L}_{N})=\left[\hat{\mathbf{x}}_{N}\right]_{1}$,
it then follows from~(\ref{eq:x1-hat-WLS}) that 
\begin{align}
\left\Vert \hat{x}_{1}(\mathcal{L}_{N+1})-\hat{x}_{1}(\mathcal{L}_{N})\right\Vert  & \leq\sum_{j=1}^{N}\left\Vert \left[\Sigma_{N+1}\right]_{1,j}-\left[\Sigma_{N}\right]_{1,j}\right\Vert \left\Vert q_{j}\right\Vert \nonumber \\
+ & \left\Vert \left[\Sigma_{N+1}\right]_{1,N+1}\right\Vert \left\Vert q_{N+1}\right\Vert .\label{eq:aux-1}
\end{align}

The main result of this section is given in Lemma~\ref{main}. It
bounds the decay rate in~(\ref{eq:aux-1}). Its proof requires a
number of lemmas, which are stated below.

We start by stating bounds for certain quantities, namely, $\mathbf{A}_{N},\mathbf{Q}_{N},\Delta_{k},\Gamma_{k}(N),\Phi_{k}(N)$.
This is done in Lemmas~\ref{lem:A}-\ref{lem:bounds}. Some of the
proofs are given in the Appendix.
\begin{lem}
\label{lem:A} For any $N\in\mathbb{N}$, 
\[
\tilde{\underline{\varepsilon}}I\leq\mathbf{A}_{N}\leq\tilde{\overline{\varepsilon}}I,
\]
with 
\begin{eqnarray*}
\tilde{\overline{\varepsilon}} & = & \max_{i}(\left\Vert \tilde{C}_{i}\right\Vert ^{2}+2\max\{\Vert\tilde{C}_{i-1,i}\Vert^{2},\Vert\tilde{C}_{i,i-1}\Vert^{2}\}\\
 &  & +2\max\{\Vert\tilde{C}_{i,i+1}\Vert^{2},\Vert\tilde{C}_{i+1,i}\Vert^{2}\})^{1/2},\\
\tilde{\underline{\varepsilon}} & = & \min_{i}\sigma_{\min}(\tilde{C}_{i}).
\end{eqnarray*}
\end{lem}
\begin{lem}
\label{lem:Q} For any $N\in\mathbb{N}$, 
\[
\tilde{\underline{q}}I\leq\mathbf{Q}_{N}\leq\tilde{\overline{q}}I,
\]
with 
\begin{align*}
\tilde{\underline{q}} & =\frac{\tilde{\underline{\varepsilon}}^{2}}{\tilde{\overline{r}}},\quad\tilde{\bar{r}}=\max_{i}\left\{ \left\Vert \tilde{R}_{i}\right\Vert ,\left\Vert \tilde{R}_{i,i+1}\right\Vert \right\} ,\\
\tilde{\overline{q}} & =\frac{\tilde{\overline{\varepsilon}}^{2}}{\tilde{\underline{r}}},\quad\tilde{\underline{r}}=\min_{i}\left\{ \sigma_{\min}(\tilde{R}_{i}),\sigma_{\min}(\tilde{R}_{i,i+1})\right\} .
\end{align*}
\end{lem}
\begin{pf}
Since $\mathbf{Q}_{N}=\mathbf{A}_{N}^{T}\mathbf{S}_{N}^{-1}\mathbf{A}_{N}$,
from Lemma~\ref{lem:A}, it follows that 
\[
\frac{\tilde{\underline{\varepsilon}}^{2}}{\sigma_{\max}(\mathbf{S}_{N})}I\leq\mathbf{Q}_{N}\leq\frac{\tilde{\overline{\varepsilon}}^{2}}{\sigma_{\min}(\mathbf{S}_{N})}I.
\]
The result then follows from $\tilde{\underline{r}}I\leq\mathbf{S}_{N}\leq\tilde{\overline{r}}I$. 
\end{pf}
\begin{lem}
\label{lem:bounds} For every $1\leq k\leq N$, 
\[
\tilde{\underline{q}}I\leq\Delta_{k},\Gamma_{k}(N),\Phi_{k}(N)\leq\tilde{\overline{q}}I.
\]
\end{lem}
Our next goal is to bound the difference $\Vert\hat{x}_{1}(\mathcal{L}_{N+1})-\hat{x}_{1}(\mathcal{L}_{N})\Vert$.
From~\eqref{guanjianyinsu} and~\eqref{eq:aux-1}, this requires
an upper bound for $\left\Vert \Phi_{j}^{-1}(N+1)-\Phi_{j}^{-1}(N)\right\Vert $.
This is given in the following lemma, whose proof appears in the Appendix.
\begin{lem}
\label{lem:decay-phi}For any $1\leq j\leq N$, we have 
\[
\left\Vert \Phi_{j}^{-1}(N+1)-\Phi_{j}^{-1}(N)\right\Vert \leq\tilde{\underline{q}}^{-1}\left(e^{\tilde{\psi}_{N}\tilde{\lambda}_{N}^{N-j}}-1\right)
\]
with 
\begin{eqnarray*}
\tilde{\psi}_{N} & = & \sqrt{\bar{n}|\mathcal{T}_{N}|}\tilde{\bar{\xi}}_{N},\\
\tilde{\bar{\xi}}_{N} & = & \max_{i\leq N}\log\sigma_{\max}[I+(\tilde{C}_{i,i+1}^{T}\tilde{R}_{i,i+1}^{-1}\tilde{C}_{i,i+1})\\
 &  & \cdot(\tilde{C}_{i}^{T}\tilde{R}_{i}^{-1}\tilde{C}_{i})^{-1}],\\
\tilde{\lambda}_{N} & = & \frac{\tilde{\alpha}_{1,N}}{\tilde{\alpha}_{1,N}+\tilde{\beta}_{1,N}}\frac{\tilde{\alpha}_{2,N}}{\tilde{\alpha}_{2,N}+\tilde{\beta}_{2,N}},
\end{eqnarray*}
where 
\begin{align*}
\tilde{\alpha}_{1,N} & =\max_{i\le N}\Vert\tilde{C}_{i,i+1}^{T}\tilde{R}_{i,i+1}^{-1}\tilde{C}_{i,i+1}\Vert,\\
\tilde{\alpha}_{2,N} & =\max_{i\le N}\Vert\tilde{C}_{i+1,i}(\tilde{C}_{i+1}^{T}\tilde{R}_{i+1}^{-1}\tilde{C}_{i+1})^{-1}\tilde{C}_{i+1,i}^{T}\Vert,\\
\tilde{\beta}_{1,N} & =\min_{i\le N}\sigma_{\min}(\tilde{C}_{i}^{T}\tilde{R}_{i}^{-1}\tilde{C}_{i}),\quad\tilde{\beta}_{2,N}=\min_{i\le N}\sigma_{\min}(\tilde{R}_{i,i+1}).
\end{align*}
\end{lem}
Combining the results in Lemmas~\ref{lem:bounds}-\ref{lem:decay-phi}
and \eqref{guanjianyinsu}, we can obtain upper bounds for $\left\Vert \left[\Sigma_{N+1}\right]_{1,j}-\left[\Sigma_{N}\right]_{1,j}\right\Vert $
and $\left\Vert \left[\Sigma_{N}\right]_{1,j}\right\Vert $. These
are given in Lemmas~\ref{lem:decay1} and \ref{lem:decay2}, respectively.
\begin{lem}
\label{lem:decay1}For any $1\leq j\leq N$, 
\[
\left\Vert \left[\Sigma_{N+1}\right]_{1,j}-\left[\Sigma_{N}\right]_{1,j}\right\Vert \leq\tilde{\overline{q}}^{-1}\tilde{r}^{j}\left(e^{\tilde{\psi}_{N}\tilde{\lambda}_{N}^{N-j}}-1\right)
\]
with $\tilde{r}=\tilde{\overline{q}}/\tilde{\underline{q}}$.
\end{lem}
\begin{pf}
From Lemma~\ref{lem:PD-bound}, for all $k\in\mathbb{N}$, 
\[
\left\Vert Q_{k,k+1}\right\Vert \leq\tilde{\overline{q}}.
\]
We then have 
\begin{align*}
 & \left\Vert \left[\Sigma_{N+1}\right]_{1,j}-\left[\Sigma_{N}\right]_{1,j}\right\Vert \\
= & \left\Vert \left(\prod_{k=1}^{j-1}\Delta_{k}^{-1}Q_{k,k+1}\right)\left(\Phi_{j}^{-1}(N+1)-\Phi_{j}^{-1}(N)\right)\right\Vert \\
\leq & (\prod_{k=1}^{j-1}\left\Vert \Delta_{k}^{-1}Q_{k,k+1}\right\Vert )\left\Vert \Phi_{j}^{-1}(N+1)-\Phi_{j}^{-1}(N)\right\Vert \\
\leq & \tilde{\overline{q}}^{j-1}(\prod_{k=1}^{j-1}\left\Vert \Delta_{k}^{-1}\right\Vert )\left\Vert \Phi_{j}^{-1}(N+1)-\Phi_{j}^{-1}(N)\right\Vert .
\end{align*}
Then, using Lemmas~\ref{lem:bounds} and \ref{lem:decay-phi}, we
get 
\begin{eqnarray*}
\left\Vert \left[\Sigma_{N+1}\right]_{1,j}-\left[\Sigma_{N}\right]_{1,j}\right\Vert  & \leq & \frac{1}{\tilde{\underline{q}}}\left(\frac{\tilde{\overline{q}}}{\tilde{\underline{q}}}\right)^{j-1}\left(e^{\tilde{\psi}_{N}\tilde{\lambda}_{N}^{N-j}}-1\right)\\
 & \leq & \tilde{\overline{q}}^{-1}\tilde{r}^{j}\left(e^{\tilde{\psi}_{N}\tilde{\lambda}_{N}^{N-j}}-1\right).
\end{eqnarray*}
\end{pf}
\begin{lem}
\label{lem:decay2} For all $1\leq j\leq N$, 
\[
\left\Vert \left[\Sigma_{N}\right]_{1,j}\right\Vert \leq\tilde{c}\tilde{\iota}^{j},
\]
with 
\begin{align*}
\tilde{c} & =\frac{\tilde{r}-1}{2\tilde{\overline{q}}\tilde{\iota}},\quad\tilde{\iota}=\frac{\sqrt{\tilde{r}}-1}{\sqrt{\tilde{r}}+1}.
\end{align*}
\end{lem}
\begin{pf}
Since $\mathbf{Q}_{N}$ is $2$-banded(please refer to Definition~\ref{defband}
in Appendix), it follows from Lemma~\ref{lem:block-banded} by letting
$a=\tilde{\underline{q}}$ and $b=\tilde{\overline{q}}$.
\end{pf}
It follows from~\eqref{eq:aux-1} that, in addition to the bounds
given in Lemmas~\ref{lem:decay1} and~\ref{lem:decay2}, we also
need an upper bound for $\Vert q_{i}\Vert$. This is given in the
following lemma.
\begin{lem}
\label{lem:qn} For any $N\in\mathbb{N}$, 
\[
\max_{n\leq N}\left\Vert q_{n}\right\Vert \leq\tilde{\eta}_{N},
\]
with 
\begin{align*}
\tilde{\eta}_{N} & =\max_{i\leq N}\frac{2^{3/2}\tilde{\overline{\varepsilon}}\sqrt{\bar{m}|\mathcal{T}_{i}|}\tilde{\bar{z}}_{N}}{\tilde{\underline{r}}},\\
\tilde{\bar{z}}_{N} & =\max_{i\leq N}\left\{ \|\tilde{z}_{i}\|_{\infty},\|\tilde{z}_{i,i+1}\|_{\infty}\right\} .
\end{align*}
\end{lem}
\begin{pf}
From~\eqref{eq:S} and \eqref{eq:q_n}, 
\begin{eqnarray}
\Vert q_{n}\Vert & \leq & \frac{\tilde{\overline{\varepsilon}}(\|y_{n}\|+\|y_{n-1}\|)}{\tilde{\underline{r}}}.\label{qnx}
\end{eqnarray}
For any $n=1,2,\ldots,N$, the result then follows from 
\begin{eqnarray*}
\Vert y_{n}\Vert & \leq & \max_{i\leq N}\sqrt{2\bar{m}|\mathcal{T}_{i}|}\Vert y_{n}\Vert_{\infty}\leq\max_{i\leq N}\sqrt{2\bar{m}|\mathcal{T}_{i}|}\tilde{\bar{z}}_{N}.
\end{eqnarray*}
\end{pf}
We now state the main result of this subsection.
\begin{lem}
\label{main} For any $1\leq J\leq N$, 
\begin{align*}
 & \left\Vert \hat{x}_{1}(\mathcal{L}_{N+1})-\hat{x}_{1}(\mathcal{L}_{N})\right\Vert \\
\leq & \tilde{\eta}_{N+1}\left(\frac{\tilde{r}^{J}}{(\tilde{r}-1)\tilde{\overline{q}}}\left(e^{\tilde{\psi}_{N}\tilde{\lambda}_{N}^{N-J}}-1\right)+\frac{2\tilde{c}}{1-\tilde{\iota}}\tilde{\iota}^{J}\right).
\end{align*}
\end{lem}
\begin{pf}
From Lemmas~\ref{lem:decay1}-\ref{lem:qn} and \eqref{eq:aux-1},
\begin{align*}
 & \left\Vert \hat{x}_{1}(\mathcal{L}_{N+1})-\hat{x}_{1}(\mathcal{L}_{N})\right\Vert \\
\leq & \sum_{j=1}^{J-1}\left\Vert \left[\Sigma_{N+1}\right]_{1,j}-\left[\Sigma_{N}\right]_{1,j}\right\Vert \left\Vert q_{j}\right\Vert \\
 & +\sum_{j=J}^{N}\left\Vert \left[\Sigma_{N+1}\right]_{1,j}-\left[\Sigma_{N}\right]_{1,j}\right\Vert \left\Vert q_{j}\right\Vert \\
 & +\left\Vert \left[\Sigma_{N+1}\right]_{1,N+1}\right\Vert \left\Vert q_{N+1}\right\Vert \\
\leq & \tilde{\eta}_{N+1}\hspace{-1mm}\left(\sum_{j=1}^{J-1}\tilde{\overline{q}}^{-1}\tilde{r}^{j}\left(e^{\tilde{\psi}_{N}\tilde{\lambda}_{N}^{N-j}}\hspace{-2mm}-1\right)\hspace{-1mm}+2\tilde{c}\sum_{j=J}^{N}\tilde{\iota}^{j}+\tilde{c}\tilde{\iota}^{N+1}\right)\\
\leq & \tilde{\eta}_{N+1}\left(\sum_{j=1}^{J-1}\tilde{\overline{q}}^{-1}\tilde{r}^{j}\left(e^{\tilde{\psi}_{N}\tilde{\lambda}_{N}^{N-J}}-1\right)+2\tilde{c}\sum_{j=J}^{N+1}\tilde{\iota}^{j}\right)\\
= & \tilde{\eta}_{N+1}\left(\tilde{\overline{q}}^{-1}\left(e^{\tilde{\psi}_{N}\tilde{\lambda}_{N}^{N-J}}-1\right)\frac{\tilde{r}^{J}-\tilde{r}}{\tilde{r}-1}+2\tilde{c}\frac{\tilde{\iota}^{J}-\tilde{\iota}^{N+2}}{1-\tilde{\iota}}\right)\\
\leq & \tilde{\eta}_{N+1}\left(\tilde{\overline{q}}^{-1}\left(e^{\tilde{\psi}_{N}\tilde{\lambda}_{N}^{N-J}}-1\right)\frac{\tilde{r}^{J}}{\tilde{r}-1}+2\tilde{c}\frac{\tilde{\iota}^{J}}{1-\tilde{\iota}}\right).
\end{align*}
\end{pf}

\subsubsection{Bound of the increment in an arbitrary graph \label{sub:est-arbitrary}}

For each $N$, the value of $\hat{x}_{1}(N)$ obtained by running
Algorithm~\ref{alg:BP} on $\mathcal{G}$ equals the one $\hat{x}_{1}(\mathcal{A}_{N})$
obtained by running the same algorithm on the equivalent acyclic graph
$\mathcal{A}_{N}$. The latter in turn equals the one $\hat{x}_{1}(\mathcal{L}_{N})$
obtained by running the algorithm on the equivalent line graph $\mathcal{L}_{N}$.
Then, from Lemma~\ref{main} we obtain the following result, which
applies to an arbitrary graph $\mathcal{G}$, and whose proof appears
in the Appendix.
\begin{lem}
\label{thm:genmain} If $\check{\kappa}<1$, then, in the notation
of Theorem~\ref{xaccuracy}, for all $N\in\mathbb{N}$, 
\[
\left\Vert \hat{x}_{1}(N+1)-\hat{x}_{1}(N)\right\Vert \leq\check{\bar{\chi}}\check{\kappa}^{N},
\]
where 
\begin{align*}
\check{\bar{\chi}} & =\frac{\check{\psi}\check{\eta}}{(\overline{q}-\underline{q})\lambda}+\frac{2\check{\eta}c}{1-\iota},\quad\check{\kappa}=\max\{\bar{u}\sqrt{\lambda},\sqrt{\bar{u}}\iota^{1/\check{\zeta}}\},
\end{align*}
with 
\begin{align*}
\check{\psi} & =\left(e^{\bar{\xi}\sqrt{\bar{n}(\bar{u}+1)}}-1\right),\quad\check{\eta}=\frac{\overline{\varepsilon}\bar{z}\sqrt{8\bar{m}(\bar{u}+1)}}{\underline{r}},\\
\check{\zeta} & =2+\log_{\frac{1}{\sqrt{\lambda}}}(\overline{q}/\underline{q}),\quad c=\frac{\overline{q}-\underline{q}}{2\overline{q}\underline{q}\iota},\\
\bar{\xi} & =\max_{j}\log\Vert I+(\sum_{k\in\mathcal{N}_{j}}C_{j,k}^{T}R_{j,k}^{-1}C_{j,k})\left(C_{j}^{T}R_{j}^{-1}C_{j}\right)^{-1}\Vert,\\
\bar{z} & =\max_{i,j}\left\{ \|z_{i}\|_{\infty},\|z_{i,j}\|_{\infty}\right\} .
\end{align*}
\end{lem}
\begin{rem}
The above lemma shows that, when $\check{\kappa}<1$, the sequence
of state estimates produced by the Algorithm~\ref{alg:BP} converges
exponentially. 
\end{rem}

\subsubsection{Accuracy analysis \label{sub:est-acc}}

Let $\hat{x}_{1}^{\mathrm{WLS}}(N)$ denote the centralized WLS estimate
of $x_{1}$ obtained by considering only the subgraph $\mathcal{N}_{1}(N-1)$
(i.e., of nodes in $\mathcal{G}$ which are within $N-1$ steps away
from node~$1$). It follows from~\cite{Taixin2013acc} that, in
an acyclic graph, if the Algorithm~\ref{alg:BP} is initialized by
$\mathbb{Q}^{0}$ (as done in~\eqref{newinitial}), it generates
the true WLS estimate. Hence, based on the definition of the {loop-free
depth} $\textit{l}_{1}$ (see Notation~\ref{l_1}), for any $N\leq\textit{l}_{1}+1$,
\begin{equation}
\hat{x}_{1}(N)=\hat{x}_{1}^{\mathrm{WLS}}(N).\label{eq:acc-x0}
\end{equation}
From Lemma~\ref{thm:genmain}, if $\check{\kappa}<1$, for any $N\in\mathbb{N}$,
\begin{equation}
\left\Vert \hat{x}_{1}(N+1)-\hat{x}_{1}(N)\right\Vert \leq\check{\bar{\chi}}\check{\kappa}^{N}.\label{eq:acc-x1}
\end{equation}

Let $\mathcal{\check{T}}_{k}$ denote the set of nodes in $\mathcal{G}$
which are precisely $k-1$ steps away from node~$1$. We collect
all the nodes in $\mathcal{\check{T}}_{k}$ and their inner connections
into a single node. This yields a graph $\mathcal{\check{L}}$ having
line topology, whose structure is given in Appendix~\ref{checkL}.
The number of nodes in $\mathcal{\check{L}}$ is given by 
\[
r_{1}=\max_{j}d_{1,j}+1,
\]
where $d_{i,j}$ is the minimum distance from node~$i$ to node~$j$
in $\mathcal{G}$.

Recall that Lemma~\ref{thm:genmain} provides a bound for the state
estimate increments for Algorithm~\ref{alg:BP}. If we follow the
steps of that proof, but considering $\mathcal{\check{L}}$ in place
of $\mathcal{L}_{N}$, we would arrive to the following result.
\begin{lem}
\label{lem:WLSmain} Recall the notations in Theorem~\ref{xaccuracy}
and Lemma~\ref{thm:genmain}, for all $N\in\mathbb{N}$, if ${\kappa}<1$,
\begin{equation}
\left\Vert \hat{x}_{1}^{\mathrm{WLS}}(N+1)-\hat{x}_{1}^{\mathrm{WLS}}(N)\right\Vert \leq\begin{cases}
{\bar{\chi}}{\kappa}^{N}, & N\leq r_{1},\\
0, & N>r_{1},
\end{cases}\label{eq:acc-x2}
\end{equation}
where 
\begin{align*}
{\bar{\chi}} & =\frac{{\bar{\psi}}{\bar{\eta}}}{(\overline{q}-\underline{q}){\omega}}+\frac{2{\bar{\eta}}{c}}{1-{\iota}}
\\
{\bar{\psi}} & =e^{{\bar{\xi}}(\bar{u}+1)\sqrt{\bar{n}}}-1,\quad{\bar{\eta}}={\overline{\varepsilon}}{\bar{z}}(\bar{u}+1)\sqrt{8\bar{m}}{\underline{r}}^{-1}.
\end{align*}
\end{lem}
Combining~\eqref{eq:acc-x0}-\eqref{eq:acc-x1} and Lemma~\ref{lem:WLSmain},
we can prove our main result.
\begin{pf}
(of Theorem~\ref{xaccuracy}) Clearly, $\check{\kappa}\leq{\kappa}<1$.
Hence, from~\eqref{eq:acc-x0}-\eqref{eq:acc-x2} 
\begin{align*}
 & \Vert\hat{x}_{1}(N)-\hat{x}_{1}^{\mathrm{WLS}}\Vert\\
= & \Vert\hat{x}_{1}(N)-\hat{x}_{1}(\textit{l}_{1}+1)+\hat{x}_{1}^{\mathrm{WLS}}(\textit{l}_{1}+1)-\hat{x}_{1}^{\mathrm{WLS}}(r_{1})\Vert\\
\leq & \Vert\hat{x}_{1}(\textit{l}_{1}+1)-\hat{x}_{1}(N)\Vert+\Vert\hat{x}_{1}^{\mathrm{WLS}}(\textit{l}_{1}+1)-\hat{x}_{1}^{\mathrm{WLS}}(r_{1})\Vert\\
\leq & \check{\bar{\chi}}\sum_{t=\textit{l}_{1}+1}^{N-1}\check{\kappa}^{t}+\bar{\chi}\sum_{t=\textit{l}_{1}+1}^{r_{1}-1}\kappa^{t}\leq\frac{\bar{\chi}}{1-\kappa}\kappa^{\textit{l}_{1}+1}+\frac{\check{\bar{\chi}}}{1-\check{\kappa}}\check{\kappa}^{\textit{l}_{1}+1}.
\end{align*}
Since $a_{1}\geq{\alpha}_{1}$, $a_{2}\geq{\alpha}_{2}$, $b_{1}\leq{\beta}_{1}$
and $b_{2}\leq{\beta}_{2}$, it follows that $\omega\geq\lambda$
and $\zeta\geq\check{\zeta}$. We also have $\check{\bar{\chi}}\leq\bar{\chi}$
and $\check{\kappa}\leq\kappa$. It then follows that 
\[
\Vert\hat{x}_{1}(N)-\hat{x}_{1}^{\mathrm{WLS}}\Vert\leq\frac{2\bar{\chi}}{1-\kappa}\kappa^{\textit{l}_{1}+1}.
\]
Since the quantity $\frac{2{\bar{\chi}}}{1-{\kappa}}$ only depends
on $\bar{m},\bar{u},\bar{n}$ and the system parameters $C_{i,j},C_{i},R_{i,j},R_{i},z_{i,j},z_{i}$
for some $i$ and $j$, the result then follows.
\end{pf}

\section{Simulations\label{sec:simulation}}

\begin{figure}[ht]
\begin{centering}
\includegraphics[width=7cm]{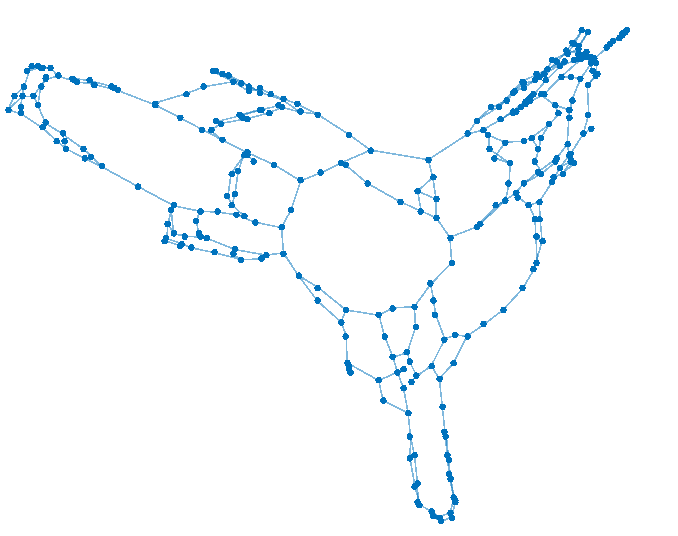} 
\par\end{centering}
\caption{Network communication graph.}
\label{graph} 
\end{figure}

In this section we present experimental evidence to support our claims,
namely, that in the case of cyclic communication graphs, the studied
distributed WLS algorithm (DWLS) converges faster than the iterative
matrix inversion (IMI) algorithm in~\cite{marelli2015distributed},
and that the accuracy of the DWLS algorithm at a given node improves
with the size of the loop-free depth of that node. To this end, we
use a network formed by 330 nodes, whose communication graph is depicted
in Figure~\ref{graph}. In this network, all nodes have the same
measurements equations, which are given by 
\begin{align*}
z_{i} & =x_{i}+v_{i};\\
z_{i,j} & =0.4x_{i}+0.4x_{j}+v_{i,j}.
\end{align*}
with $x_{i}\in\mathbb{R}^{3}$, $R_{i}=R_{i,j}=0.01$, for all $i=1,2,\ldots,330$
and $j\in\mathcal{N}_{i}$.

\begin{figure}[ht]
\begin{centering}
\includegraphics[width=8cm]{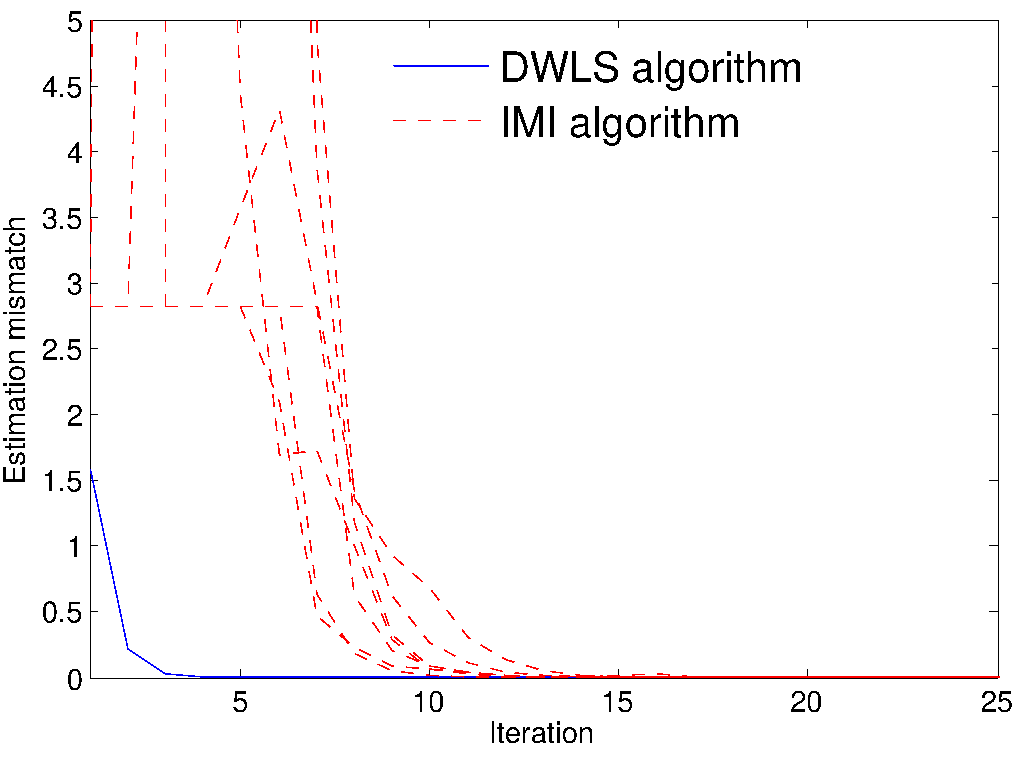} 
\par\end{centering}
\caption{Comparison between DWLS and IMI algorithms. The delayed start $\delta_{\text{IMI}}$
of the IMI algorithm ranges from $0$ to $6$.}
\label{comparison} 
\end{figure}

In the first simulation we compare the convergence rate of the DWLS
and IMI methods. As explained in~\cite{marelli2015distributed},
before starting with the matrix inversion iterations, the IMI method
needs to invest a number $\delta_{\text{IMI}}$ of iterations in order
to obtain estimates of the largest and smallest eigenvalues of certain
matrix. This delayed start is required in order to avoid that the
transients caused by too rough estimates of these eigenvalues brings
the estimation mismatch (with respect to the estimation yield by the
centralized WLS method) to very big values from which the algorithm
would take a long time to converge. In Figure~\ref{comparison} we
show the combined estimation mismatch of all 330 nodes, yield by the
DWLS algorithm and the IMI algorithm with $\delta_{\text{IMI}}$ ranging
from $0$ to $6$. We see that the DWLS algorithm converges much faster
than the IMI one, regardless of the value of $\delta_{\text{IMI}}$
used in the latter.

\begin{figure}[ht]
\begin{centering}
\includegraphics[width=7cm]{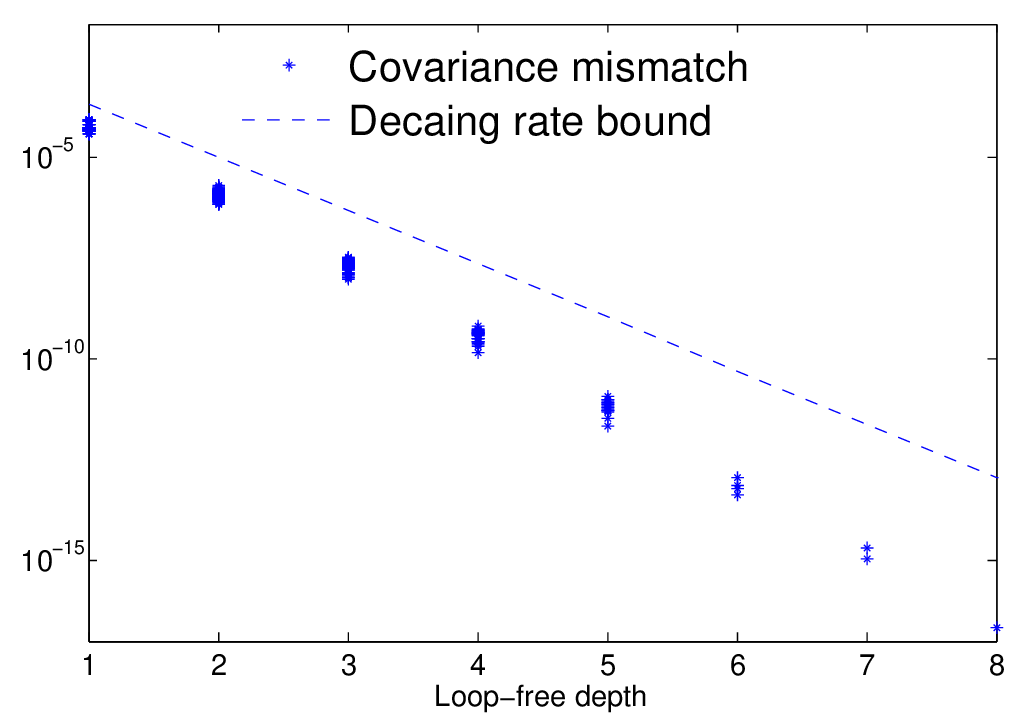} 
\par\end{centering}
\caption{Covariance mismatch between DWLS and centralized WLS.}
\label{accuracy_cov} 
\end{figure}

\begin{figure}[ht]
\begin{centering}
\includegraphics[width=7cm]{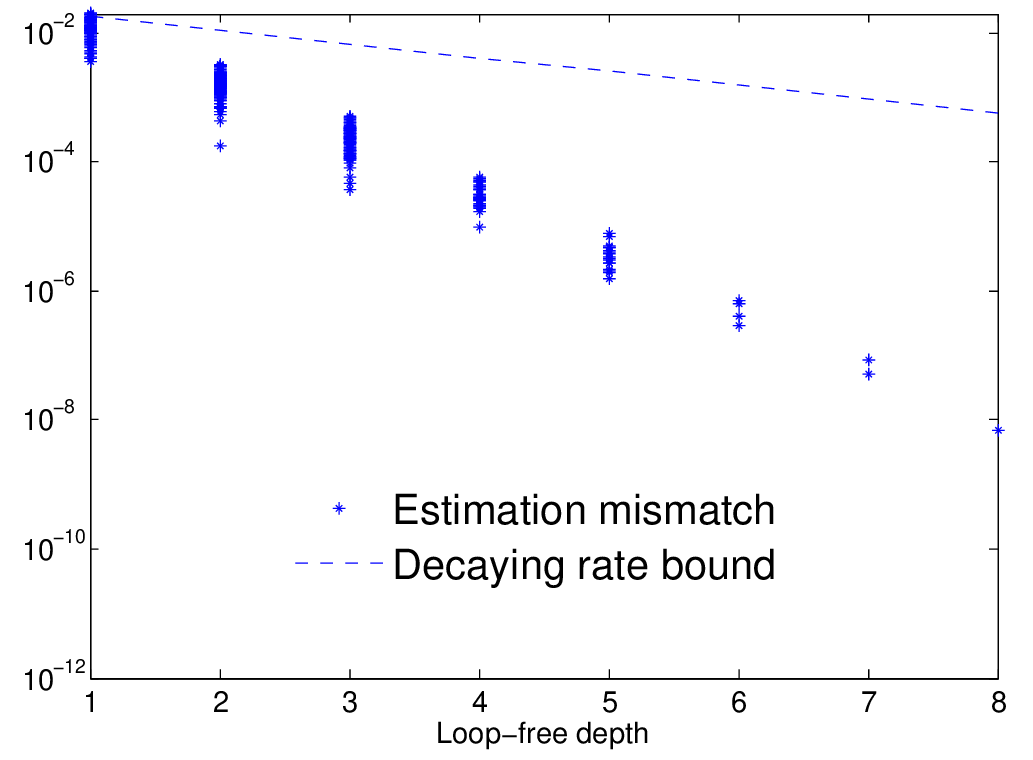} 
\par\end{centering}
\caption{Estimation mismatch between DWLS and centralized WLS.}
\label{accuracy_err} 
\end{figure}

In the second simulation we evaluate the mismatches between the estimation
and its associated covariance, produced at each node, and at time
$l_{i}+1$ (recall that $l_{i}$ denotes the loop-free depth of node
$i$), with respect to those yield by the centralized WLS method.
Figures~\ref{accuracy_cov} and \ref{accuracy_err} show these differences
for the covariance and estimate, respectively, for each node, as a
function of the loop-free depth. We see how both differences decay
exponentially with the loop-free depths of each node. We also show
in the same figures the bound on these decays derived in Theorems~\ref{disLSE}
and~\ref{xaccuracy}, respectively.

\section{Conclusions\label{sec:conclusion}}

A recently proposed distributed WLS estimation algorithm converges
in finite time if the communication graph is acyclic. We studied the
accuracy of this algorithm, when used in cyclic graphs. We showed
that, for a class of system satisfying certain requirements in terms
of topological sparsity and signal-to-noise ratio, the error between
the state estimate yield by this distributed algorithm, and that from
centralized WLS, decrease exponentially at each node, with the increase
of its local loop-free depth. The same property holds for the difference
between the estimation error covariance produced by the distributed
algorithm and that from centralized WLS. The derived expressions are
explicit and easy to interpret. An implication of our results is that,
even in applications where the communication graph has a cyclic topology,
due to its faster convergence, the studied algorithm may be a preferred
option over algorithms based on iterative matrix inversion, provided
that the loop-free depths of those nodes of interest are sufficiently
large.

\appendix

\section{Proof of Proposition~\ref{prop:riemannian}}

Properties~1)-2) come from definition and P.~947 in \cite{bougerol1993kalman}.

Property~3): Let $X\in\mathbb{R}^{n\times n}$ and $\Gamma_{k}$
denote the set of all $k$-dimensional subspaces of $\mathbb{R}^{n}$.
Recall that $\sigma_{1}(X)\geq\cdots\geq\sigma_{n}(X)$ denote the
singular-values of $X$. We have 
\begin{eqnarray}
\sigma_{k}\left(PQ^{-1}\right) & = & \sigma_{k}\left(Q^{-1/2}PQ^{-1/2}\right)\nonumber \\
 & = & \max_{\mathcal{V}\in\Gamma_{k}}\min_{x\in\mathcal{V}}\frac{\left\langle Q^{-1/2}PQ^{-1/2}x,x\right\rangle }{\left\langle x,x\right\rangle }\nonumber \\
 & = & \max_{\mathcal{V}\in\Gamma_{k}}\min_{x\in\mathcal{V}}\frac{\left\langle PQ^{-1/2}x,Q^{-1/2}x\right\rangle }{\left\langle x,x\right\rangle }\nonumber \\
 & = & \max_{\mathcal{V}\in\Gamma_{k}}\min_{x\in\mathcal{V}}\frac{\left\langle Px,x\right\rangle }{\left\langle Q^{1/2}x,Q^{1/2}x\right\rangle }\nonumber \\
 & = & \max_{\mathcal{V}\in\Gamma_{k}}\min_{x\in\mathcal{V}}\frac{\left\langle Px,x\right\rangle }{\left\langle Qx,x\right\rangle }.\label{prop3}
\end{eqnarray}
For a given subspace $\mathcal{V}\in\Gamma_{k}$, let 
\[
\Xi(\mathcal{V})=\min_{x\in\mathcal{V}}\frac{\left\langle Px,x\right\rangle }{\left\langle Qx,x\right\rangle }.
\]
Let also $\tilde{P}=BPB^{T}$ and $\tilde{Q}=BQB^{T}$, we then have
\begin{eqnarray*}
\sigma_{k}\left(\tilde{P}\tilde{Q}^{-1}\right) & = & \max_{\mathcal{V}\in\Gamma_{k}}\min_{x\in\mathcal{V}}\frac{\left\langle \tilde{P}x,x\right\rangle }{\left\langle \tilde{Q}x,x\right\rangle }\\
 & = & \max_{\mathcal{V}\in\Gamma_{k}}\min_{x\in\mathcal{V}}\frac{\left\langle PB^{T}x,B^{T}x\right\rangle }{\left\langle QB^{T}x,B^{T}x\right\rangle }\\
 & = & \max_{\mathcal{V}\in\Gamma_{k}}\Xi\left(B^{T}\mathcal{V}\right)\\
 & \leq & \max_{\mathcal{W}\in\Gamma_{k}}\Xi\left(\mathcal{W}\right)\\
 & = & \sigma_{k}\left(PQ^{-1}\right).
\end{eqnarray*}
Then, Property~3) follows because the number of non-zero eigenvalues
of $PQ^{-1}$ is no less than that of $\tilde{P}\tilde{Q}^{-1}$.

Property~4): For any vector $x$, we have 
\begin{eqnarray*}
 &  & \left\langle Q^{-1/2}\left(P+W\right)Q^{-1/2}x,x\right\rangle \\
 & = & \left\langle Q^{-1/2}PQ^{-1/2}x,x\right\rangle +\left\langle Q^{-1/2}WQ^{-1/2}x,x\right\rangle \\
 & \geq & \left\langle PQ^{-1}x,x\right\rangle .
\end{eqnarray*}
Hence, $\delta(P+W,Q)\ge\delta(P,Q)$ holds from above inequality,
equality \eqref{prop3} and the fact that $\sigma_{k}(PQ^{-1})\ge1$
for all $k$.

Property~5): The case when $B$ is invertible follows immediately
from the invariance of eigenvalues under similarity transformations\cite{bougerol1993kalman}.
Hence, we show the case when $B$ is non-invertible. Let $\Lambda=Q^{1/2}P^{-1}Q^{1/2}$
and decompose $B$ into $B=U^{T}\mbox{diag}\{\Delta,0\}V$, where
$U$ and $V$ are unitary matrices and $\Delta>0$ is diagonal. Let
also $VQ^{-1/2}=[T^{T}\ X^{T}]^{T}$, with $T$ having as many rows
as $\Delta$. Then, 
\begin{align*}
BP^{-1}B^{T} & =U^{T}\begin{bmatrix}\Delta & 0\\
0 & 0
\end{bmatrix}VQ^{-1/2}\Lambda Q^{-1/2}V^{T}\begin{bmatrix}\Delta^{T} & 0\\
0 & 0
\end{bmatrix}U\\
 & =U^{T}\begin{bmatrix}\Delta & 0\\
0 & 0
\end{bmatrix}\begin{bmatrix}T\\
X
\end{bmatrix}\Lambda\begin{bmatrix}T^{T} & X^{T}\end{bmatrix}\begin{bmatrix}\Delta^{T} & 0\\
0 & 0
\end{bmatrix}U\\
 & =U^{T}\begin{bmatrix}\Delta T\Lambda T^{T}\Delta^{T} & 0\\
0 & 0
\end{bmatrix}U.
\end{align*}
Similarly, we obtain 
\[
BQ^{-1}B^{T}=U^{T}\begin{bmatrix}\Delta TT^{T}\Delta^{T} & 0\\
0 & 0
\end{bmatrix}U.
\]
Thus, for any $\varepsilon>0$, it follows that 
\begin{align}
 & \delta(\varepsilon I+BP^{-1}B^{T},\varepsilon I+BQ^{-1}B^{T})\nonumber \\
= & \delta(\varepsilon I+\Delta T\Lambda T^{T}\Delta^{T},\varepsilon I+\Delta TT^{T}\Delta^{T}).\label{eq:aux-2}
\end{align}
Since $\Delta T$ has full row rank, it follows from Property~3)
that 
\begin{align}
 & \delta(\Delta T\Lambda T^{T}\Delta^{T},\Delta TT^{T}\Delta^{T})\nonumber \\
\leq & \delta(\Lambda,I)=\delta(P^{-1},Q^{-1})=\delta(P,Q).\label{eq:aux-3}
\end{align}
Then, from~\eqref{eq:aux-2} and \eqref{eq:aux-3}, we have 
\begin{align*}
 & \delta(W+BP^{-1}B^{T},W+BQ^{-1}B^{T})\\
= & \lim_{\varepsilon\rightarrow0^{+}}\delta[(W-\varepsilon I)+(\varepsilon I+BP^{-1}B^{T}),\\
 & (W-\varepsilon I)+(\varepsilon I+BQ^{-1}B^{T})]\\
\overset{\text{(b)}}{\le} & \frac{\alpha}{\alpha+\beta}\lim_{\varepsilon\rightarrow0^{+}}\delta(\varepsilon I+BP^{-1}B^{T},\varepsilon I+BQ^{-1}B^{T})\\
= & \frac{\alpha}{\alpha+\beta}\lim_{\varepsilon\rightarrow0^{+}}\delta(\varepsilon I+\Delta T_{1}\Lambda T_{1}^{T}\Delta^{T},\varepsilon I+\Delta T_{1}T_{1}^{T}\Delta^{T})\\
= & \frac{\alpha}{\alpha+\beta}\delta(\Delta T_{1}\Lambda T_{1}^{T}\Delta^{T},\Delta T_{1}T_{1}^{T}\Delta^{T})\\
\leq & \frac{\alpha}{\alpha+\beta}\delta(P,Q),
\end{align*}
where (b) above follows from \cite[Proposition 1.6]{bougerol1993kalman}
and 
\begin{align*}
\alpha & =\lim_{\varepsilon\rightarrow0^{+}}\max\{\|\varepsilon I+BP^{-1}B^{T}\|,\|\varepsilon I+BQ^{-1}B^{T}\|\}\\
 & =\max\{\|BP^{-1}B^{T}\|,\|BQ^{-1}B^{T}\|\},\\
\beta & =\lim_{\varepsilon\rightarrow0^{+}}\sigma_{\min}(W-\varepsilon I)=\sigma_{\min}(W).
\end{align*}

Property~6): We have 
\[
\delta(P,Q)=\sqrt{\sum_{k=1}^{n}\log^{2}\sigma_{k}(PQ^{-1})}\geq\log\left\Vert PQ^{-1}\right\Vert .
\]
Then, 
\begin{align*}
\left\Vert PQ^{-1}\right\Vert  & \leq e^{\delta(P,Q)}\Rightarrow PQ^{-1}\leq e^{\delta(P,Q)}I\Rightarrow\\
P & \leq e^{\delta(P,Q)}Q\Rightarrow P-Q\leq\left(e^{\delta(P,Q)}-1\right)Q,
\end{align*}
and the result follows.

\section{Proof of Lemma~\ref{lem:decay-Q-1}}

For each $N\in\mathbb{N}$, consider the line $\mathcal{L}_{N}$ built
from $\mathcal{G}$ as described in Section~\ref{sub:conv-line-2}.
Recall the Notation~\ref{initial}, let $\tilde{Q}_{i}(n,\mathbb{Q})$,
$n\leq N$, denote the information matrix at Node~$i$ and step $n$,
when running the distributed WLS algorithm in the network $\mathcal{L}_{N}$
with initial condition set $\mathbb{Q}$. And the similar notation
follows $\tilde{Q}_{i\rightarrow i-1}(n,\mathbb{Q})$ and $\tilde{Q}_{i\rightarrow i+1}(n,\mathbb{Q})$.
To simplify the expressions, we denote $\tilde{Q}_{i\rightarrow i-1,i}(n,\mathbb{Q}):=Q_{i}(n,\mathbb{Q})-Q_{i-1\rightarrow i}(n-1,\mathbb{Q})$
in this proof.

Then, based on the properties in Proposition~\ref{prop:riemannian},
the Riemannian distance between $\tilde{Q}_{1}(N,\mathbb{Q}_{1})$
and $\tilde{Q}_{1}(N,\mathbb{Q}_{2})$ is given by 
\begin{align*}
 & \delta\left(\tilde{Q}_{1}(N,\mathbb{Q}_{1}),\tilde{Q}_{1}(N,\mathbb{Q}_{2})\right)\\
= & \delta(\tilde{C}_{1}^{T}\tilde{R}_{1}^{-1}\tilde{C}_{1}+\tilde{C}_{1,2}^{T}(\tilde{R}_{1,2}+\tilde{C}_{2,1}\tilde{Q}_{2\rightarrow1,2}^{-1}(N-1,\mathbb{Q}_{1})\\
 & \cdot\tilde{C}_{2,1}^{T})^{-1}\tilde{C}_{1,2},\tilde{C}_{1}^{T}\tilde{R}_{1}^{-1}\tilde{C}_{1}+\tilde{C}_{1,2}^{T}(\tilde{R}_{1,2}\\
 & +\tilde{C}_{2,1}\tilde{Q}_{2\rightarrow1,2}^{-1}(N-1,\mathbb{Q}_{2})\tilde{C}_{2,1}^{T})^{-1}\tilde{C}_{1,2})\\
\leq & \frac{\tilde{\alpha}_{1,N}}{\tilde{\alpha}_{1,N}+\tilde{\beta}_{1,N}}\delta(\tilde{R}_{1,2}+\tilde{C}_{2,1}\tilde{Q}_{2\rightarrow1,2}^{-1}(N-1,\mathbb{Q}_{1})\tilde{C}_{2,1}^{T}\\
 & ,\tilde{R}_{1,2}+\tilde{C}_{2,1}\tilde{Q}_{2\rightarrow1,2}^{-1}(N-1,\mathbb{Q}_{2})\tilde{C}_{2,1}^{T})\\
\leq & \frac{\tilde{\alpha}_{1,N}}{\tilde{\alpha}_{1,N}+\tilde{\beta}_{1,N}}\frac{\tilde{\alpha}_{2,N}}{\tilde{\alpha}_{2,N}+\tilde{\beta}_{2,N}}\\
 & \cdot\delta(\tilde{Q}_{2\rightarrow1,2}(N-1,\mathbb{Q}_{1}),\tilde{Q}_{2\rightarrow1,2}(N-1,\mathbb{Q}_{2})),
\end{align*}
where 
\begin{align*}
\tilde{\alpha}_{1,N} & =\max_{i\le N}\Vert\tilde{C}_{i,i+1}^{T}\tilde{R}_{i,i+1}^{-1}\tilde{C}_{i,i+1}\Vert,\\
\tilde{\alpha}_{2,N} & =\max_{i\le N}\Vert\tilde{C}_{i+1,i}(\tilde{C}_{i+1}^{T}\tilde{R}_{i+1}^{-1}\tilde{C}_{i+1})^{-1}\tilde{C}_{i+1,i}^{T}\Vert,\\
\tilde{\beta}_{1,N} & =\min_{i\le N}\sigma_{\min}(\tilde{C}_{i}^{T}\tilde{R}_{i}^{-1}\tilde{C}_{i}),\\
\tilde{\beta}_{2,N} & =\min_{i\le N}\sigma_{\min}(\tilde{R}_{i,i+1}).
\end{align*}
Recall the definition of line $\mathcal{L}_{N}$ in Subsection~\ref{sub:conv-line-2},
it follows that $\tilde{\alpha}_{1,N}\leq\alpha_{1}$, $\tilde{\alpha}_{2,N}\leq\alpha_{2}$
and $\tilde{\beta}_{1,N}\geq\beta_{1}$, $\tilde{\beta}_{2,N}\geq\beta_{2}$.
Then, we have 
\begin{align*}
 & \delta\left(\tilde{Q}_{1}(N,\mathbb{Q}_{1}),\tilde{Q}_{1}(N,\mathbb{Q}_{2})\right)\\
\leq & \lambda\delta\left(\tilde{Q}_{2\rightarrow1,2}(N-1,\mathbb{Q}_{1}),\tilde{Q}_{2\rightarrow1,2}(N-1,\mathbb{Q}_{2})\right)\\
\leq & \lambda^{N-1}\delta\left(\tilde{Q}_{N\rightarrow N-1,N}(1,\mathbb{Q}_{1}),\tilde{Q}_{N\rightarrow N-1,N}(1,\mathbb{Q}_{2})\right).
\end{align*}

Now, for any initial condition $\mathbb{Q}$, 
\[
\tilde{Q}_{N\rightarrow N-1,N}(1,\mathbb{Q})=\tilde{C}_{N}^{T}\tilde{R}_{N}^{-1}\tilde{C}_{N}+\tilde{Q}_{N+1\rightarrow N}(0,\mathbb{Q}),
\]
hence 
\begin{eqnarray*}
\tilde{C}_{N}^{T}\tilde{R}_{N}^{-1}\tilde{C}_{N} & \leq & \tilde{Q}_{N\rightarrow N-1,N}(1,\mathbb{Q})\\
 & \leq & \tilde{C}_{N}^{T}\tilde{R}_{N}^{-1}\tilde{C}_{N}+\tilde{C}_{N,N+1}^{T}\tilde{R}_{N,N+1}^{-1}\tilde{C}_{N,N+1}.
\end{eqnarray*}
Recall that $|\mathcal{T}_{i}|$ is the number of nodes in the $i$-th
layer of $\mathcal{A}_{N}$. Since $|\mathcal{T}_{i}|\leq(\bar{u}+1)\bar{u}^{i-1}$,
we then have 
\begin{align*}
 & \delta\left(\tilde{Q}_{N\rightarrow N-1,N}(1,\mathbb{Q}_{1}),\tilde{Q}_{N\rightarrow N-1,N}(1,\mathbb{Q}_{2})\right)\\
\leq & \delta(\tilde{C}_{N}^{T}\tilde{R}_{N}^{-1}\tilde{C}_{N}\\
 & ,\tilde{C}_{N}^{T}\tilde{R}_{N}^{-1}\tilde{C}_{N}+\tilde{C}_{N,N+1}^{T}\tilde{R}_{N,N+1}^{-1}\tilde{C}_{N,N+1})\\
\leq & \sqrt{\bar{n}|\mathcal{T}_{N}|}\\
 & \max_{i\leq N}\log\Vert I+(\tilde{C}_{i,i+1}^{T}\tilde{R}_{i,i+1}^{-1}\tilde{C}_{i,i+1})(\tilde{C}_{i}^{T}\tilde{R}_{i}^{-1}\tilde{C}_{i})^{-1}\Vert\\
\leq & \sqrt{\bar{n}}\sqrt{|\mathcal{T}_{N}|}\\
 & \max_{j}\log\left\Vert I+\left(\sum_{k\in\mathcal{N}_{j}}C_{j,k}^{T}R_{j,k}^{-1}C_{j,k}\right)\left(C_{j}^{T}R_{j}^{-1}C_{j}\right)^{-1}\right\Vert \\
\leq & \bar{\delta}\bar{u}^{(N-1)/2}.
\end{align*}
Thus, 
\begin{multline*}
\delta\left(Q_{1}(N,\mathbb{Q}_{1}),Q_{1}(N,\mathbb{Q}_{2})\right)=\delta\left(\tilde{Q}_{1}(N,\mathbb{Q}_{1}),\tilde{Q}_{1}(N,\mathbb{Q}_{2})\right)\\
\leq\bar{\delta}\bar{u}^{(N-1)/2}\lambda_{1}^{N-1}=\bar{\delta}\rho^{N-1}.
\end{multline*}

\section{Proof of Lemma~\ref{lem:accu-Q}}

For each $N\in\mathbb{N}$, consider the network $\mathcal{L}_{N}$
built from $\mathcal{G}$ as described in Section~\ref{sub:conv-line-2}.
Recall the notations $\tilde{Q}_{i}(N,\mathbb{Q})$ and $\tilde{Q}_{i\rightarrow i-1,i}(N,\mathbb{Q})$
for any initial condition $\mathbb{Q}$ in the proof of Lemma~\ref{lem:decay-Q-1},
We have that 
\begin{multline*}
\tilde{Q}_{i\rightarrow i-1,i}(N,\mathbb{Q})=\tilde{C}_{i}^{T}\tilde{R}_{i}^{-1}\tilde{C}_{i}+\tilde{C}_{i,i+1}^{T}(\tilde{R}_{i,i+1}\\
+\tilde{C}_{i,i+1}(\tilde{Q}_{i+1\rightarrow i,i+1}(N-1,\mathbb{Q}))^{-1}\tilde{C}_{i+1,i}^{T})^{-1}\tilde{C}_{i,i+1}
\end{multline*}
is a increasing function of $\tilde{Q}_{i+1\rightarrow i,i+1}(N-1,\mathbb{Q})$.
Also, $\tilde{Q}_{j\rightarrow j-1,j}(2,\mathbb{Q}^{\mathrm{M}})\leq\tilde{Q}_{j\rightarrow j-1,j}(1,\mathbb{Q}^{\mathrm{M}})$
and $\tilde{Q}_{j\rightarrow j-1,j}(2,\mathbb{Q}^{0})\geq\tilde{Q}_{j\rightarrow j-1,j}(1,\mathbb{Q}^{0})$,
for any $j$. Hence, $\tilde{Q}_{i\rightarrow i-1,i}(N,\mathbb{Q}^{\mathrm{M}})$
is a decreasing function of $N$ and $\tilde{Q}_{i\rightarrow i-1,i}(N,\mathbb{Q}^{0})$
is an increasing function of $N$. Thus, we have 
\begin{equation}
Q_{1}^{-1}(\textit{l}_{1}+1,\mathbb{Q}^{\mathrm{M}})\leq Q_{1}^{-1}(N)\leq Q_{1}^{-1}(\textit{l}_{1}+1,\mathbb{Q}^{0}).\label{eq:aux1}
\end{equation}

The computation of $\tilde{Q}_{1}^{-1}(\textit{l}_{1}+1,\mathbb{Q}^{0})$
is done by implicitly assuming that $\tilde{Q}_{\textit{l}_{1}+2\rightarrow\textit{l}_{1}+1}(0,\mathbb{Q}^{0})=0$.
On the other hand, notice that $\tilde{Q}_{1}^{-1}(\textit{l}_{1}+1,\mathbb{Q}^{0})$
would equal $\mathrm{Cov}_{1}^{\mathrm{WLS}}$ if $\tilde{Q}_{\textit{l}_{1}+2\rightarrow\textit{l}_{1}+1}(0,\mathbb{Q}^{0})$
is given with a properly chosen positive semi-definite value. Hence,
it follows that $\tilde{Q}_{1}^{-1}(\textit{l}_{1}+2,\mathbb{Q}^{0})\geq\mathrm{Cov}_{1}^{\mathrm{WLS}}$.

Similarly, on the computation of $\tilde{Q}_{1}^{-1}(\textit{l}_{1}+1,\mathbb{Q}^{\mathrm{M}})$,
it is implicitly assumed that $\tilde{Q}_{\textit{l}_{1}+2\rightarrow\textit{l}_{1}+1}(0,\mathbb{Q}^{\mathrm{M}})=\tilde{C}_{\textit{l}_{1}+1,\textit{l}_{1}+2}^{T}\tilde{R}_{\textit{l}_{1}+1,\textit{l}_{1}+2}^{-1}\tilde{C}_{\textit{l}_{1}+1,\textit{l}_{1}+2}$,
which means that the estimate of $\tilde{x}_{\textit{l}_{1}+2}$ do
not have error, i.e., $\hat{\tilde{x}}_{\textit{l}_{1}+2}=\tilde{x}_{\textit{l}_{1}+2}$.
Hence, $\tilde{Q}_{1}^{-1}(\textit{l}_{1}+1,\mathbb{Q}^{\mathrm{M}})\leq\mathrm{Cov}_{1}^{\mathrm{WLS}}$,
and we have 
\begin{equation}
Q_{1}^{-1}(\textit{l}_{1}+1,\mathbb{Q}^{\mathrm{M}})\leq\mathrm{Cov}_{1}^{\mathrm{WLS}}\leq Q_{1}^{-1}(\textit{l}_{1}+1,\mathbb{Q}^{0}).\label{eq:aux2}
\end{equation}
Then, the result follows from~(\ref{eq:aux1}) and~(\ref{eq:aux2}).

\section{Proof of Lemma~\ref{lem:A}}

Recall that $\mathbf{x}_{N}^{T}=\left[\tilde{x}_{1}^{T},\cdots,\tilde{x}_{N}^{T}\right]$,
we have 
\begin{align*}
 & \Vert\mathbf{A}_{N}\mathbf{x}_{N}\Vert^{2}\\
= & \sum_{i=1}^{N}\left\Vert \tilde{C}_{i}\tilde{x}_{i}\right\Vert ^{2}+\sum_{i=1}^{N-1}\left\Vert \tilde{C}_{i,i+1}\tilde{x}_{i}+\tilde{C}_{i+1,i}\tilde{x}_{i+1}\right\Vert ^{2}\\
\leq & \sum_{i=1}^{N}\left\Vert \tilde{C}_{i}\right\Vert ^{2}\left\Vert \tilde{x}_{i}\right\Vert ^{2}+\sum_{i=1}^{N-1}\left\Vert [\tilde{C}_{i,i+1},\tilde{C}_{i+1,i}]\right\Vert ^{2}\left\Vert \left[\begin{array}{c}
\tilde{x}_{i}\\
\tilde{x}_{i+1}
\end{array}\right]\right\Vert ^{2}
\end{align*}
\begin{align*}
= & \sum_{i=1}^{N}\left\Vert \tilde{C}_{i}\right\Vert ^{2}\left\Vert \tilde{x}_{i}\right\Vert ^{2}+\sum_{i=1}^{N-1}\left\Vert [\tilde{C}_{i,i+1},\tilde{C}_{i+1,i}]\right\Vert ^{2}\left\Vert \tilde{x}_{i}\right\Vert ^{2}\\
 & +\sum_{i=2}^{N}\left\Vert [\tilde{C}_{i-1,i},\tilde{C}_{i,i-1}]\right\Vert ^{2}\left\Vert \tilde{x}_{i}\right\Vert ^{2}\\
\leq & \sum_{i=1}^{N}\left(\left\Vert \tilde{C}_{i}\right\Vert ^{2}+\left\Vert [\tilde{C}_{i,i+1},\tilde{C}_{i+1,i}]\right\Vert ^{2}\right.\\
 & \left.+\left\Vert [\tilde{C}_{i-1,i},\tilde{C}_{i,i-1}]\right\Vert ^{2}\right)\left\Vert \tilde{x}_{i}\right\Vert ^{2}\\
\leq & \max_{i}\{\Vert\tilde{C}_{i}\Vert^{2}+\Vert[\tilde{C}_{i-1,i},\tilde{C}_{i,i-1}]\Vert^{2}\\
 & +\Vert[\tilde{C}_{i,i+1},\tilde{C}_{i+1,i}]\Vert^{2}\}\sum_{j=1}^{N}\left\Vert \tilde{x}_{j}\right\Vert ^{2}.
\end{align*}
Combining with 
\[
\Vert[\tilde{C}_{i,i+1},\tilde{C}_{i+1,i}]\Vert\leq\sqrt{2}\max\{\Vert\tilde{C}_{i,i+1}\Vert,\Vert\tilde{C}_{i+1,i}\Vert\},
\]
we have 
\[
\Vert\mathbf{A}_{N}\mathbf{x}_{N}\Vert^{2}\leq\tilde{\overline{\varepsilon}}^{2}\left\Vert \mathbf{x}_{N}\right\Vert ^{2}.
\]

Also, 
\begin{align*}
\Vert\mathbf{A}_{N}\mathbf{x}_{N}\Vert^{2} & \geq\sigma_{\min}^{2}(\tilde{C}_{1})\|\tilde{x}_{1}\|^{2}+\sigma_{\min}^{2}(\tilde{C}_{2})\|\tilde{x}_{2}\|^{2}+\ldots\\
 & +\sigma_{\min}^{2}(\tilde{C}_{N})\|\tilde{x}_{N}\|^{2}\geq\tilde{\underline{\varepsilon}}^{2}\|\mathbf{x}_{N}\|^{2},
\end{align*}
completing the proof.

\section{Proof of Lemma~\ref{lem:bounds}}

From~\cite[S 4]{marelli2015distributed}, it follows that $\Delta_{k}$
is the information matrix at node $k$ of the system $\mathbf{y}_{k}=\mathbf{A}_{k}\mathbf{x}_{k}+\mathbf{w}_{k}$.
Hence, $\Delta_{k}^{-1}=\left[\Sigma_{k}\right]_{kk}$. Then, 
\begin{multline*}
\sigma_{\min}\left(\Delta_{k}\right)=\sigma_{\max}^{-1}\left(\Delta_{k}^{-1}\right)=\sigma_{\max}^{-1}\left(\left[\Sigma_{k}\right]_{kk}\right)\\
\overset{\mathrm{(a)}}{\geq}\sigma_{\max}^{-1}\left(\Sigma_{k}\right)=\sigma_{\min}\left(\mathbf{Q}_{k}\right)\geq\tilde{\underline{q}},
\end{multline*}
where (a) follows from~\cite[S 2]{zhang2006schur}. Following a similar
argument, we obtain 
\[
\sigma_{\max}\left(\Delta_{k}\right)\leq\tilde{\overline{q}},
\]
which proves the result for $\Delta_{k}$. The result for $\Gamma_{k}(N)$
follows from a similar argument, after noting that $\Gamma_{k}(N)$
is the information matrix at node $k$ of the system 
\[
\left[\begin{array}{c}
y_{k}\\
\vdots\\
y_{N}
\end{array}\right]=\left[\begin{array}{ccc}
A_{kk} & A_{k,k+1} & 0\\
0 & \ddots & 0\\
\vdots & \ddots & A_{N-1,N}\\
0 & \cdots & A_{N,N}
\end{array}\right]\left[\begin{array}{c}
\tilde{x}_{k}\\
\vdots\\
\tilde{x}_{N}
\end{array}\right]+\left[\begin{array}{c}
w_{k}\\
\vdots\\
w_{N}
\end{array}\right].
\]
Finally, $\Phi_{k}(N)$ is the information matrix at node $k$ of
the system $\mathbf{y}_{N}=\mathbf{A}_{N}\mathbf{x}_{N}+\mathbf{w}_{N}$,
and the result for $\Phi_{k}(N)$ also follows.

\section{Proof of Lemma~\ref{lem:decay-phi}}

We split the proof into four steps:

Step 1: From~\eqref{eq:S} and~\eqref{eq:Q-2}, we obtain 
\begin{equation}
Q_{jj}=\begin{cases}
\tilde{C}_{1}^{T}\tilde{R}_{1}^{-1}\tilde{C}_{1}+\tilde{C}_{1,2}^{T}\tilde{R}_{1,2}^{-1}\tilde{C}_{1,2}, & j=1,\\
\tilde{C}_{j}^{T}\tilde{R}_{j}^{-1}\tilde{C}_{j}+\tilde{C}_{j,j+1}^{T}\tilde{R}_{j,j+1}^{-1}\tilde{C}_{j,j+1}\\
+\tilde{C}_{j,j-1}^{T}\tilde{R}_{j,j-1}^{-1}\tilde{C}_{j,j-1}, & 2\leq j\leq N-1,\\
\tilde{C}_{N}^{T}\tilde{R}_{N}^{-1}\tilde{C}_{N}\\
+\tilde{C}_{N,N-1}^{T}\tilde{R}_{N-1,N}^{-1}\tilde{C}_{N,N-1}, & j=N
\end{cases}\label{eq:Q1}
\end{equation}
and 
\begin{equation}
Q_{j,j+1}=\tilde{C}_{j,j+1}^{T}\tilde{R}_{j,j+1}^{-1}\tilde{C}_{j+1,j}.\label{eq:Q2}
\end{equation}
Let 
\[
\tilde{\Gamma}_{j}(N)=\Gamma_{j}(N)-\tilde{C}_{j,j-1}^{T}\tilde{R}_{j-1,j}^{-1}\tilde{C}_{j,j-1}.
\]
Putting~\eqref{eq:Q1}-\eqref{eq:Q2} into~\eqref{eq:Phi}, for
any $j=1,2,\ldots,N-1$, we get 
\begin{eqnarray}
\tilde{\Gamma}_{j}(N) & = & \tilde{C}_{j}^{T}\tilde{R}_{j}^{-1}\tilde{C}_{j}+\tilde{C}_{j,j+1}^{T}\check{\Gamma}_{j}^{-1}(N)\tilde{C}_{j,j+1}.\label{eq:Gamma1}
\end{eqnarray}
with 
\begin{align*}
\check{\Gamma}_{j}^{-1}(N) & =\tilde{R}_{j,j+1}^{-1}-\tilde{R}_{j,j+1}^{-1}\tilde{C}_{j+1,j}\Gamma_{j+1}^{-1}(N)\tilde{C}_{j+1,j}^{T}\tilde{R}_{j,j+1}^{-1}\\
 & =\tilde{R}_{j,j+1}^{-1}-\tilde{R}_{j,j+1}^{-1}\tilde{C}_{j+1,j}[\tilde{\Gamma}_{j+1}(N)\\
 & +\tilde{C}_{j+1,j}^{T}\tilde{R}_{j,j+1}^{-1}\tilde{C}_{j+1,j}]^{-1}\tilde{C}_{j+1,j}^{T}\tilde{R}_{j,j+1}^{-1}.
\end{align*}
Using the matrix inversion lemma, for any $j=1,2,\ldots,N-1$, we
obtain 
\begin{equation}
\check{\Gamma}_{j}(N)=\tilde{R}_{j,j+1}+\tilde{C}_{j+1,j}\tilde{\Gamma}_{j+1}^{-1}(N)\tilde{C}_{j+1,j}^{T}.\label{eq:Gamma2}
\end{equation}
Following similar steps, we also define 
\[
\tilde{\Delta}_{j}=\Delta_{j}-\tilde{C}_{j,j+1}^{T}\tilde{R}_{j,j+1}^{-1}\tilde{C}_{j,j+1},
\]
and obtain that 
\begin{align*}
\tilde{\Delta}_{j} & =\tilde{C}_{j}^{T}\tilde{R}_{j}^{-1}\tilde{C}_{j}+\tilde{C}_{j,j-1}^{T}\check{\Delta}_{j}^{-1}\tilde{C}_{j,j-1},\\
\check{\Delta}_{j} & =\tilde{R}_{j-1,j}+\tilde{C}_{j-1,j}\tilde{\Delta}_{j-1}^{-1}\tilde{C}_{j-1,j}^{T}.
\end{align*}

Step 2: From~\eqref{eq:Gamma1}-\eqref{eq:Gamma2}, and using the
properties of Riemannian Distance in Proposition~\ref{prop:riemannian},
we get 
\begin{align*}
 & \delta\left(\tilde{\Gamma}_{j}(N+1),\tilde{\Gamma}_{j}(N)\right)\\
= & \delta(\tilde{C}_{j}^{T}\tilde{R}_{j}^{-1}\tilde{C}_{j}+\tilde{C}_{j,j+1}^{T}\check{\Gamma}_{j}^{-1}(N+1)\tilde{C}_{j,j+1}\\
 & ,\tilde{C}_{j}^{T}\tilde{R}_{j}^{-1}\tilde{C}_{j}+\tilde{C}_{j,j+1}^{T}\check{\Gamma}_{j}^{-1}(N)\tilde{C}_{j,j+1})\\
\leq & \frac{\tilde{\pi}_{1,j}(N)}{\tilde{\pi}_{1,j}(N)+\tilde{\tau}_{1,j}(N)}\delta\left(\check{\Gamma}_{j}^{-1}(N+1),\check{\Gamma}_{j}^{-1}(N)\right)\\
= & \frac{\tilde{\pi}_{1,j}(N)}{\tilde{\pi}_{1,j}(N)+\tilde{\tau}_{1,j}(N)}\delta\left(\check{\Gamma}_{j}(N+1),\check{\Gamma}_{j}(N)\right)\\
= & \frac{\tilde{\pi}_{1,j}(N)}{\tilde{\pi}_{1,j}(N)+\tilde{\tau}_{1,j}(N)}\delta(\tilde{R}_{j,j+1}+\tilde{C}_{j+1,j}\tilde{\Gamma}_{j+1}^{-1}(N+1)\\
 & \cdot\tilde{C}_{j+1,j}^{T},\tilde{R}_{j,j+1}+\tilde{C}_{j+1,j}\tilde{\Gamma}_{j+1}^{-1}(N)\tilde{C}_{j+1,j}^{T})\\
\leq & \frac{\tilde{\pi}_{1,j}(N)}{\tilde{\pi}_{1,j}(N)+\tilde{\tau}_{1,j}(N)}\frac{\tilde{\pi}_{2,j}(N)}{\tilde{\pi}_{2,j}(N)+\tilde{\tau}_{2,j}(N)}\\
 & \cdot\delta(\tilde{\Gamma}_{j+1}^{-1}(N+1),\tilde{\Gamma}_{j+1}^{-1}(N))\\
= & \frac{\tilde{\pi}_{1,j}(N)}{\tilde{\pi}_{1,j}(N)+\tilde{\tau}_{1,j}(N)}\frac{\tilde{\pi}_{2,j}(N)}{\tilde{\pi}_{2,j}(N)+\tilde{\tau}_{2,j}(N)}\\
 & \cdot\delta\left(\tilde{\Gamma}_{j+1}(N+1),\tilde{\Gamma}_{j+1}(N)\right),
\end{align*}
where 
\begin{eqnarray*}
\tilde{\pi}_{1,j}(N) & = & \max_{\tilde{N}=N,N-1}\left\Vert \tilde{C}_{j,j+1}^{T}\check{\Gamma}_{j}^{-1}(\tilde{N})\tilde{C}_{j,j+1}\right\Vert \\
 & \leq & \max_{i\leq N}\Vert\tilde{C}_{i,i+1}^{T}\tilde{R}_{i,i+1}^{-1}\tilde{C}_{i,i+1}\Vert=\tilde{\alpha}_{1,N},\\
\tilde{\tau}_{1,j}(N) & = & \sigma_{\min}\left(\tilde{C}_{j}^{T}\tilde{R}_{j}^{-1}\tilde{C}_{j}\right)\\
 & \geq & \min_{i\leq N}\sigma_{\min}\left(\tilde{C}_{i}^{T}\tilde{R}_{i}^{-1}\tilde{C}_{i}\right)=\tilde{\beta}_{1,N},
\end{eqnarray*}
\begin{eqnarray*}
\tilde{\pi}_{2,j}(N) & = & \max_{\tilde{N}=N,N-1}\Vert\tilde{C}_{j+1,j}\tilde{\Gamma}_{j+1}^{-1}(\tilde{N})\tilde{C}_{j+1,j}^{T}\Vert\\
 & \leq & \max_{i\leq N}\Vert\tilde{C}_{i+1,i}(\tilde{C}_{i}^{T}\tilde{R}_{i}^{-1}\tilde{C}_{i})^{-1}\tilde{C}_{i+1,i}^{T}\Vert\\
 & = & \tilde{\alpha}_{2,N},\\
\tilde{\tau}_{2,j}(N) & = & \sigma_{\min}\left(\tilde{R}_{j,j+1}\right)\geq\min_{i\leq N}\sigma_{\min}\left(\tilde{R}_{i,i+1}\right)\\
 & = & \tilde{\beta}_{2,N}.
\end{eqnarray*}
Recall that $\tilde{\lambda}_{N}=\frac{\tilde{\alpha}_{1,N}}{\tilde{\alpha}_{1,N}+\tilde{\beta}_{1,N}}\frac{\tilde{\alpha}_{2,N}}{\tilde{\alpha}_{2,N}+\tilde{\beta}_{2,N}}$,
we then have 
\begin{align*}
 & \delta\left(\tilde{\Gamma}_{j}(N+1),\tilde{\Gamma}_{j}(N)\right)\\
\leq & \tilde{\lambda}_{N}\delta\left(\tilde{\Gamma}_{j+1}(N+1),\tilde{\Gamma}_{j+1}(N)\right)\\
\leq & \tilde{\lambda}_{N}^{N-j}\delta\left(\tilde{\Gamma}_{N}(N+1),\tilde{\Gamma}_{N}(N)\right)\\
\leq & \tilde{\lambda}_{N}^{N-j}\delta\left(\tilde{C}_{N}^{T}\tilde{R}_{N}^{-1}\tilde{C}_{N},\tilde{C}_{N}^{T}\tilde{R}_{N}^{-1}\tilde{C}_{N}+\tilde{C}_{N,N+1}^{T}\tilde{R}_{N,N+1}^{-1}\tilde{C}_{N,N+1}\right)\\
 & \leq\tilde{\lambda}_{N}^{N-j}\sqrt{\bar{n}|\mathcal{T}_{N}|\tilde{\bar{\xi}}_{N}^{2}}=\tilde{\psi}_{N}\tilde{\lambda}_{N}^{N-j},
\end{align*}

Step 3: From~\eqref{eq:Phi}, we have 
\begin{align*}
\Phi_{j}(N) & =\Gamma_{j}(N)-Q_{j,j-1}\Delta_{j-1}^{-1}Q_{j-1,j}\\
 & =\Gamma_{j}(N)+\Delta_{j}-Q_{j,j}\\
 & =\Gamma_{j}(N)+\Delta_{j}-\tilde{C}_{j}^{T}\tilde{R}_{j}^{-1}\tilde{C}_{j}\\
 & -\tilde{C}_{j,j+1}^{T}\tilde{R}_{j,j+1}^{-1}\tilde{C}_{j,j+1}-\tilde{C}_{j,j-1}^{T}\tilde{R}_{j,j-1}^{-1}\tilde{C}_{j,j-1}\\
 & =\tilde{\Gamma}_{j}(N)+\tilde{\Delta}_{j}-\tilde{C}_{j}^{T}\tilde{R}_{j}^{-1}\tilde{C}_{j}\\
 & =\tilde{\Gamma}_{j}(N)+\tilde{C}_{j,j-1}^{T}\check{\Delta}_{j}^{-1}\tilde{C}_{j,j-1}.
\end{align*}
Then, 
\begin{multline*}
\delta\left(\Phi_{j}^{-1}(N+1),\Phi_{j}^{-1}(N)\right)=\delta\left(\Phi_{j}(N+1),\Phi_{j}(N)\right)\\
=\delta(\tilde{\Gamma}_{j}(N+1)+\tilde{C}_{j,j-1}^{T}\check{\Delta}_{j}^{-1}\tilde{C}_{j,j-1},\tilde{\Gamma}_{j}(N)+\tilde{C}_{j,j-1}^{T}\check{\Delta}_{j}^{-1}\tilde{C}_{j,j-1})\\
\leq\delta\left(\tilde{\Gamma}_{j}(N+1),\tilde{\Gamma}_{j}(N)\right)\leq\tilde{\psi}_{N}\tilde{\lambda}_{N}^{N-j}.
\end{multline*}

Step 4: From the Proposition~\ref{prop:riemannian}, 
\begin{multline*}
\left\Vert \Phi_{j}^{-1}(N+1)-\Phi_{j}^{-1}(N)\right\Vert \\
\leq\left(e^{\delta\left(\Phi_{j}^{-1}(N+1),\Phi_{j}^{-1}(N)\right)}-1\right)\left\Vert \Phi_{j}^{-1}(N)\right\Vert .
\end{multline*}
Now, using Lemma~\ref{lem:bounds}, we get 
\begin{multline*}
\left\Vert \Phi_{j}^{-1}(N+1)-\Phi_{j}^{-1}(N)\right\Vert \\
\leq\left(e^{\tilde{\psi}_{N}\tilde{\lambda}_{N}^{N-j}}-1\right)\left\Vert \Phi_{j}^{-1}(N)\right\Vert \leq\tilde{\underline{q}}^{-1}\left(e^{\tilde{\psi}_{N}\tilde{\lambda}_{N}^{N-j}}-1\right).
\end{multline*}

\section{Proof of Lemma~\ref{thm:genmain}}

Following the equivalence between $\hat{x}_{1}(N)$ in $\mathcal{G}$
and $\hat{x}_{1}(\mathcal{L}_{N})$ in $\mathcal{L}_{N}$, we have
\[
\left\Vert \hat{x}_{1}(N+1)-\hat{x}_{1}(N)\right\Vert =\left\Vert \hat{x}_{1}(\mathcal{L}_{N+1})-\hat{x}_{1}(\mathcal{L}_{N})\right\Vert .
\]
Since $\tilde{\psi}_{N}=\sqrt{\bar{n}|\mathcal{T}_{N}|}\tilde{\bar{\xi}}_{N}$
and $\tilde{\eta}_{N+1}=\max_{i\leq N+1}\frac{2^{3/2}\tilde{\overline{\varepsilon}}\sqrt{\bar{m}|\mathcal{T}_{i}|}\tilde{\bar{z}}_{N}}{\tilde{\underline{r}}}$,
from Lemma~\ref{main} and taking $J=\lceil\frac{N}{\check{\alpha}}\rceil$,
we have 
\begin{align*}
 & \left\Vert \hat{x}_{1}(\mathcal{L}_{N+1})-\hat{x}_{1}(\mathcal{L}_{N})\right\Vert \\
\leq & \tilde{\eta}_{N+1}\left(\frac{\tilde{r}^{N/\check{\alpha}+1}}{(\tilde{r}-1)\tilde{\overline{q}}}\left(e^{\tilde{\psi}_{N}\tilde{\lambda}_{N}^{N-N/\check{\alpha}-1}}-1\right)+\frac{2\tilde{c}}{1-\tilde{\iota}}\tilde{\iota}^{N/\check{\alpha}}\right)\\
= & \max_{i\leq N+1}\frac{2^{3/2}\tilde{\overline{\varepsilon}}\sqrt{\bar{m}|\mathcal{T}_{i}|}\tilde{\bar{z}}_{N}}{\tilde{\underline{r}}}[\frac{\tilde{r}^{N/\check{\alpha}+1}}{(\tilde{r}-1)\tilde{\overline{q}}}\\
 & \cdot\left(e^{\sqrt{\bar{n}|\mathcal{T}_{N}|}\tilde{\bar{\xi}}_{N}\tilde{\lambda}_{N}^{N-N/\check{\alpha}-1}}-1\right)+\frac{2\tilde{c}}{1-\tilde{\iota}}\tilde{\iota}^{N/\check{\alpha}}]
\end{align*}
Following the similar definition as in Lemma~\ref{lem:decay1}, we
let $r=\overline{q}/\underline{q}$. Substituting the structure of
$\mathcal{L}_{N}$ in Subsection~\ref{sub:conv-line-2}, it follows
that 
\begin{multline}
\left\Vert \hat{x}_{1}(\mathcal{L}_{N+1})-\hat{x}_{1}(\mathcal{L}_{N})\right\Vert \\
\leq\max_{i\leq N+1}\frac{2^{3/2}{\overline{\varepsilon}}\sqrt{\bar{m}|\mathcal{T}_{i}|}\bar{z}}{{\underline{r}}}[\frac{{r}^{(N/\check{\zeta})+1}}{({r}-1){\overline{q}}}\\
\cdot\left(e^{\sqrt{\bar{n}|\mathcal{T}_{N}|}{\bar{\xi}}\lambda^{N-(N/\check{\zeta})-1}}-1\right)+\frac{2{c}}{1-\iota}\iota^{N/\check{\zeta}}].\label{xinguanjian}
\end{multline}
Combining with $\max_{i\leq N}|\mathcal{T}_{i}|\leq(\bar{u}+1)\bar{u}^{N-1}$,
the inequality in \eqref{xinguanjian} is upper bounded by 
\begin{multline}
\left\Vert \hat{x}_{1}(\mathcal{L}_{N+1})-\hat{x}_{1}(\mathcal{L}_{N})\right\Vert \\
\leq\frac{2^{3/2}{\overline{\varepsilon}}\sqrt{\bar{m}(\bar{u}+1)\bar{u}^{N}}\bar{z}}{{\underline{r}}}[\frac{{r}^{(N/\check{\zeta})+1}}{({r}-1){\overline{q}}}\\
\cdot\left(e^{\sqrt{\bar{n}(\bar{u}+1)\bar{u}^{N-1}}{\bar{\xi}}\lambda^{N-(N/\check{\zeta})-1}}-1\right)+\frac{2{c}}{1-\iota}\iota^{N/\check{\zeta}}].\label{xianzhi}
\end{multline}

Under condition $\check{\kappa}<1$ and $J=\lceil\frac{N}{\check{\zeta}}\rceil$,
combining that $\check{\zeta}>2$, it follows that 
\[
\bar{u}^{(N-1)/2}{\lambda}^{N-(N/\check{\zeta})-1}\leq({\lambda}\bar{u})^{(N-1)/2}\leq\check{\kappa}<1.
\]

Then, following the Lemma~\ref{lem:exp-bound}, it follows 
\begin{multline*}
e^{\sqrt{\bar{n}(\bar{u}+1)\bar{u}^{N-1}}{\bar{\xi}}\lambda^{N-(N/\check{\zeta})-1}}-1\\
\leq\left(e^{\bar{\xi}\sqrt{\bar{n}(\bar{u}+1)}}-1\right)(\sqrt{\bar{u}})^{N}\lambda^{N-(N/\check{\zeta})-1}.
\end{multline*}

Denote ${d}={r}/{\lambda}$, it follows that 
\[
\check{\zeta}=\log_{\frac{1}{\sqrt{\lambda}}}d,
\]
and we have 
\begin{align*}
 & \left\Vert \hat{x}_{1}(N+1)-\hat{x}_{1}(N)\right\Vert \\
\leq & \frac{2^{3/2}{\overline{\varepsilon}}\sqrt{\bar{m}(\bar{u}+1)\bar{u}^{N}}\bar{z}}{{\underline{r}}}(\frac{\left(e^{\bar{\xi}\sqrt{\bar{n}(\bar{u}+1)}}-1\right){d}^{N/{\check{\zeta}}+1}}{({r}-1){\overline{q}}}\\
 & \cdot\left(\sqrt{\bar{u}}{\lambda}\right)^{N}+\frac{2{c}}{1-\iota}\iota^{N/{\check{\zeta}}})\\
\leq & \frac{{\overline{\varepsilon}}{\bar{z}}\sqrt{8\bar{m}(\bar{u}+1)}}{{\underline{r}}}\\
 & \cdot\left(\frac{\left(e^{{\bar{\xi}}\sqrt{\bar{n}(\bar{u}+1)}}-1\right){d}}{({r}-1){\overline{q}}}\left(\bar{u}\sqrt{{\lambda}}\right)^{N}\hspace{-3mm}+\frac{2{c}}{1-\iota}(\sqrt{\bar{u}}\iota^{1/\check{\zeta}})^{N}\right)\\
\leq & \check{\eta}\left(\frac{\check{\psi}{d}}{({r}-1){\overline{q}}}\left(\bar{u}\sqrt{{\lambda}}\right)^{N}+\frac{2{c}}{1-\iota}(\sqrt{\bar{u}}\iota^{1/\check{\zeta}})^{N}\right)\\
= & \frac{\check{\psi}{r}\check{\eta}}{({r}-1)\lambda{\overline{q}}}(\bar{u}\sqrt{{\lambda}})^{N}+\frac{2\check{\eta}{c}}{1-\iota}(\sqrt{\bar{u}}\iota^{1/\check{\zeta}})^{N}\leq\check{\bar{\chi}}\check{\kappa}^{N}.
\end{align*}

\section{The structure of $\mathcal{\check{L}}$ in Subsection~\ref{sub:est-acc}}

\label{checkL}

Suppose that the $j\in\mathcal{\check{T}}_{k}$, denote the parent
set of Node~$j$ by $\mathcal{\check{P}}_{j}:=\{p|p\in\mathcal{\check{T}}_{k-1},p\in\mathcal{N}_{j}\}$
and child set of Node~$j$ by $\mathcal{\check{S}}_{j}:=\{p|p\in\mathcal{\check{T}}_{k+1},p\in\mathcal{N}_{j}\}$.
For any $i\in\mathcal{\check{T}}_{k}$, its rank number in $\mathcal{\check{T}}_{k}$
is denoted by $b_{\mathcal{\check{T}}_{k}}(i)$. Furthermore, for
any $i,j\in\mathcal{\check{T}}_{k}$, if $j\in\mathcal{N}_{i}$, the
rank number of connection $\{i,j\}$ is denoted by $(i,j)$(a scalar
number). And the set of all connections between nodes in $\mathcal{\check{T}}_{k}$
is denoted by $\mathbb{U}_{\mathcal{\check{T}}_{k}}=\{(i,j)|i,j\in\mathcal{\check{T}}_{k}~\text{and}~j\in\mathcal{N}_{i}\}$.

For each $n=1,\cdots,r_{1}$, the $n$-th Node in $\mathcal{\check{L}}$
for $\hat{x}_{1}^{WLS}(k)$ follows 
\begin{eqnarray}
\check{z}_{n} & = & \check{C}_{n}\check{x}_{n}+\check{v}_{n},\label{eq:lineWLS1}\\
\check{z}_{n,n+1} & = & \check{C}_{n,n+1}\check{x}_{n}+\check{C}_{n+1,n}\check{x}_{n+1}+\check{v}_{n,n+1},\label{eq:lineWLS2}
\end{eqnarray}
where 
\begin{eqnarray*}
\check{x}_{n} & = & \begin{bmatrix}{x}_{\mathcal{\check{T}}_{n}^{T}(1)}, & {x}_{\mathcal{\check{T}}_{n}^{T}(2)}, & \ldots, & {x}_{\mathcal{\check{T}}_{n}(|\mathcal{\check{T}}_{n}|)}^{T}\end{bmatrix}^{T}.
\end{eqnarray*}
Here, 
\begin{eqnarray*}
\check{C}_{n} & = & \begin{bmatrix}\check{H}_{\mathcal{\check{T}}_{n}}\\
\check{V}_{\mathcal{\check{T}}_{n}}
\end{bmatrix},\\
\check{C}_{n,n+1} & = & \text{diag}\{F_{1,\mathcal{\check{T}}_{n}(1)},F_{1,\mathcal{\check{T}}_{n}(2)},\ldots,F_{1,\mathcal{\check{T}}_{n}(|\mathcal{\check{T}}_{n}|)}\},\\
\check{C}_{n+1,n} & = & \text{diag}\{F_{2,\mathcal{\check{T}}_{n}(1)},F_{2,\mathcal{\check{T}}_{n}(2)},\ldots,F_{2,\mathcal{\check{T}}_{n}(|\mathcal{\check{T}}_{n}|)}\}
\end{eqnarray*}
with 
\begin{eqnarray*}
\check{H}_{\mathcal{\check{T}}_{n}} & = & \text{diag}\{C_{\mathcal{\check{T}}_{n}(1)},C_{\mathcal{\check{T}}_{n}(2)},\ldots,C_{\mathcal{\check{T}}_{n}(|\mathcal{\check{T}}_{n}|)}\},\\
\left[\check{V}_{\mathcal{\check{T}}_{n}}\right]_{x,y} & = & \begin{cases}
C_{u,v}, & x=(u,v)\text{ and }y=b_{\mathcal{\check{T}}_{n}}(u),\\
C_{v,u}, & x=(u,v)\text{ and }y=b_{\mathcal{\check{T}}_{n}}(v),\\
0, & \text{else.}
\end{cases}
\end{eqnarray*}
as the $(x,y)$-th block of $\check{V}_{\mathcal{\check{T}}_{n}}$,
and 
\begin{eqnarray*}
F_{1,i} & = & \begin{bmatrix}C_{i,\mathcal{\check{S}}_{i}(1)}^{T} & C_{i,\mathcal{\check{S}}_{i}(2)}^{T} & \ldots & C_{i,\mathcal{\check{S}}_{i}(|\mathcal{\check{S}}_{i}|)}^{T}\end{bmatrix}^{T},\\
F_{2,i} & = & \begin{bmatrix}C_{\mathcal{\check{S}}_{i}(1),i}^{T} & C_{\mathcal{\check{S}}_{i}(2),i}^{T} & \ldots & C_{\mathcal{\check{S}}_{i}(|\mathcal{\check{S}}_{i}|),i}^{T}\end{bmatrix}^{T}.
\end{eqnarray*}

Also, $\check{v}_{n}\sim\mathcal{N}\left(0,\check{R}_{n}\right)$
and $\check{v}_{n,n+1}\sim\mathcal{N}\left(0,\check{R}_{n,n+1}\right)$,
with 
\begin{eqnarray*}
\check{R}_{n} & = & \mathrm{diag}\{R_{\mathcal{\check{T}}_{n}(1)},R_{\mathcal{\check{T}}_{n}(2)},\ldots,R_{\mathcal{\check{T}}_{n}(|\mathcal{\check{T}}_{n}|)}\\
 &  & ,R_{u,v}|_{(u,v)=1},\ldots,R_{u,v}|_{(u,v)=|\mathbb{U}_{\mathcal{\check{T}}_{n}}|}\},
\end{eqnarray*}
and 
\begin{eqnarray*}
\check{R}_{n,n+1} & = & \mathrm{diag}\{F_{3,\mathcal{\check{T}}_{n}(1)},F_{3,\mathcal{\check{T}}_{n}(2)},\ldots,F_{3,\mathcal{\check{T}}_{n}(|\mathcal{\check{T}}_{n}|)}\},\\
F_{3,i} & = & \text{diag}\{R_{i,\mathcal{\check{S}}_{i}(1)},R_{i,\mathcal{\check{S}}_{i}(2)},\ldots,R_{i,\mathcal{\check{S}}_{i}(|\mathcal{\check{S}}_{i}|)}\}.
\end{eqnarray*}
Finally 
\begin{eqnarray*}
\check{z}_{n} & = & \text{col}\{z_{\mathcal{\check{T}}_{n}(1)},z_{\mathcal{\check{T}}_{n}(2)},\ldots,z_{\mathcal{\check{T}}_{n}(|\mathcal{\check{T}}_{n}|)}\\
 &  & ,z_{u,v}|_{(u,v)=1},\ldots,z_{u,v}|_{(u,v)=|\mathbb{U}_{\mathcal{\check{T}}_{n}}|}\},
\end{eqnarray*}
and 
\begin{eqnarray*}
\check{z}_{n,n+1} & = & \begin{bmatrix}F_{4,\mathcal{\check{T}}_{n}(1)}^{T}, & F_{4,\mathcal{\check{T}}_{n}(2)}^{T}, & \ldots, & F_{4,\mathcal{\check{T}}_{n}(|\mathcal{\check{T}}_{n}|)}^{T}\end{bmatrix}^{T},\\
F_{4,i} & = & \begin{bmatrix}z_{i,\mathcal{\check{S}}_{i}(1)}^{T}, & z_{i,\mathcal{\check{S}}_{i}(2)}^{T}, & \ldots, & z_{i,\mathcal{\check{S}}_{i}(|\mathcal{\check{S}}_{i}|)}^{T}\end{bmatrix}^{T}.
\end{eqnarray*}
Note that the line $\mathcal{\check{L}}$ for $\hat{x}_{1}^{WLS}(k)$
given in \eqref{eq:lineWLS1} and \eqref{eq:lineWLS2} is another
expression of origin graph $\mathcal{G}$.

\section{Some Additional lemmas}
\begin{lem}
\label{lem:exp-bound} For every $x\in\mathbb{R}$ and $0\leq y\leq1$,
\[
e^{xy}-1\leq\left(e^{x}-1\right)y.
\]
\end{lem}
\begin{pf}
Fix $x\in\mathbb{R}$. Let $f_{x}(y)=e^{xy}-1$ and $g_{x}(y)=\left(e^{x}-1\right)y$.
We have that 
\begin{align*}
f_{x}(0) & =0=g_{x}(0),\quad f_{x}(1)=e^{x}-1=g_{x}(1).
\end{align*}
Hence, the result follows since the function $f_{x}$ is concave and
$g_{x}$ is linear.
\end{pf}
\begin{lem}
\label{lem:PD-bound} If $\left[\begin{array}{cc}
A & B^{T}\\
B & C
\end{array}\right]\geq0$, then $\left\Vert B\right\Vert \leq\sqrt{\left\Vert A\right\Vert \left\Vert C\right\Vert }$. 
\end{lem}
\begin{pf}
Taking the Schur complement of $A$, we have that 
\[
A-B^{T}C^{-1}B\geq0.
\]
Hence, 
\begin{multline*}
\left\Vert A\right\Vert \geq\left\Vert B^{T}C^{-1}B\right\Vert =\max_{\left\Vert x\right\Vert =1}x^{T}B^{T}C^{-1}Bx\\
\geq\sigma_{\min}\left(C^{-1}\right)\max_{\left\Vert x\right\Vert =1}x^{T}B^{T}Bx=\frac{\left\Vert B\right\Vert ^{2}}{\left\Vert C\right\Vert },
\end{multline*}
and the result follows.
\end{pf}
\begin{defn}
\label{defband} We say that the block matrix $A=\left[A_{ij}\right]$
is $m$-banded if $A_{ij}=0$, whenever $\left|i-j\right|>m/2.$ 
\end{defn}
The following lemma is a straightforward generalization of~\cite[Proposition 2.2]{demko1984decay}
to the case of block matrices.
\begin{lem}
\label{lem:block-banded}If $A$ is an $m$-banded block matrix, for
any $(i,j)$-th block of $A^{-1}$, we have 
\[
\left\Vert \left[A^{-1}\right]_{ij}\right\Vert \leq c\lambda^{\left|i-j\right|}
\]
where 
\begin{align*}
c & =\frac{r-1}{2b},\quad\lambda=\left(\frac{\sqrt{r}-1}{\sqrt{r}+1}\right)^{\frac{2}{m}},
\end{align*}
$r=b/a$ and $a\leq\sigma(A)\leq b$. 
\end{lem}
\begin{pf}
Fix $i$ and $j$ and let $n<\frac{2}{m}\left|i-j\right|$. From~\cite[Proposition 2.1]{demko1984decay},
there exists a polynomial $p$, of degree $n$, satisfying 
\[
\sup_{x\in\left[a,b\right]}\left|\frac{1}{x}-p(x)\right|=Kq^{n+1},
\]
with 
\begin{align*}
K & =\frac{\left(1+\sqrt{r}\right)^{2}}{2ar},\quad q=\frac{\sqrt{r}-1}{\sqrt{r}+1}.
\end{align*}
Then, 
\begin{align*}
\left\Vert A^{-1}-p(A)\right\Vert  & =\sup_{x\in\sigma(A)}\left|\frac{1}{x}-p(x)\right|\leq Kq^{n+1}.
\end{align*}
Since $A^{n}$ is $nm$-banded, it follows that $p(A)$ is $nm$-banded.
Hence, 
\[
|i-j|>\frac{mn}{2}\Rightarrow\left[p(A)\right]_{ij}=0.
\]
Then 
\begin{align*}
\left\Vert \left[A^{-1}\right]_{ij}\right\Vert  & =\left\Vert \left[A^{-1}-p(A)\right]_{ij}\right\Vert \leq\left\Vert A^{-1}-p(A)\right\Vert \\
 & \leq Kq^{n+1}\leq Kq^{\frac{2}{m}\left|i-j\right|+1}=Kq\left(q^{\frac{2}{m}}\right)^{\left|i-j\right|},
\end{align*}
and the result follows.
\end{pf}
\bibliographystyle{plain}
\bibliography{mybibf}

\end{document}